\documentclass[aps,pra, twocolumn,superscriptaddress,floatfix,nobibnotes]{revtex4-1}
\usepackage{amsmath,amsfonts,amssymb,bbold}
\usepackage{graphicx}
\usepackage{epstopdf}
\usepackage[english]{babel}
\usepackage{natbib}
\usepackage{hyperref}
\usepackage[utf8]{inputenc}

\usepackage{letltxmacro}

\LetLtxMacro{\ORIGselectlanguage}{\selectlanguage}
\makeatletter
\DeclareRobustCommand{\selectlanguage}[1]{%
  \@ifundefined{alias@\string#1}
    {\ORIGselectlanguage{#1}}
    {\begingroup\edef\x{\endgroup
       \noexpand\ORIGselectlanguage{\@nameuse{alias@#1}}}\x}%
}
\newcommand{\definelanguagealias}[2]{%
  \@namedef{alias@#1}{#2}%
}
\makeatother

\definelanguagealias{en}{english}
\definelanguagealias{eng}{english}
\definelanguagealias{English}{english}

\newcommand{\ket}[1]{\ensuremath{\left|#1\right\rangle}}

\newcommand{\<}{\langle}
\renewcommand{\>}{\rangle}
\providecommand{\abs}[1]{\left\lvert#1\right\rvert}
\providecommand{\norm}[1]{\lVert#1\rVert}

\providecommand{\tr}{{\rm tr}}
\renewcommand{\phi}{\varphi}

\begin{document}
\title{Statistical signatures of multimode single-photon added and subtracted states of light}

\author{Mattia Walschaers} 
\email{mattia.walschaers@lkb.upmc.fr}
\affiliation{Laboratoire Kastler Brossel, UPMC-Sorbonne Universit\'es, CNRS, ENS-PSL Research University, Coll\`ege de France, CNRS; 4 place Jussieu, F-75252 Paris, France}
\author{Claude Fabre}
\affiliation{Laboratoire Kastler Brossel, UPMC-Sorbonne Universit\'es, CNRS, ENS-PSL Research University, Coll\`ege de France, CNRS; 4 place Jussieu, F-75252 Paris, France}
\author{Valentina Parigi}
\affiliation{Laboratoire Kastler Brossel, UPMC-Sorbonne Universit\'es, CNRS, ENS-PSL Research University, Coll\`ege de France, CNRS; 4 place Jussieu, F-75252 Paris, France}
\author{Nicolas Treps}
\affiliation{Laboratoire Kastler Brossel, UPMC-Sorbonne Universit\'es, CNRS, ENS-PSL Research University, Coll\`ege de France, CNRS; 4 place Jussieu, F-75252 Paris, France}

\date{\today}

\begin{abstract}
The addition or subtraction of a photon from a Gaussian state of light is a versatile and experimentally feasible procedure to create non-Gaussian states. In multimode setups, these states manifest a wide range of phenomena when the photon is added or subtracted in a mode-tunable way. In this contribution, we derive the truncated correlations, which are multimode generalisations of cumulants, between quadratures in different modes as statistical signatures of these states. These correlations are then used to obtain the full multimode Wigner function, the properties of which are subsequently studied. In particular we investigate the effect of impurity in the subtraction or addition process, and evaluate its impact on the negativity of the Wigner function. Finally, we elaborate on the generation of inherent entanglement through subtraction or addition of a photon from a pure squeezed vacuum.
\end{abstract}

\maketitle

\section{Introduction}\label{sec:Intro}

Continuous variable (CV) quantum optics has ample advantages for quantum information processing. The most notable strength of general optical systems is their resilience against decoherence, which proves useful for quantum protocols. 
%The CV regime, moreover, is of particular interest for its entanglement properties. 
%In CV setups, the degree of freedom which are entangled, are the system's modes (in optics these are normalised solutions to Maxwell's equations). 
In CV quantum optics, states with arbitrary many entangled modes (normalised solutions to Maxwell's equations) can be deterministically generated \cite{PhysRevLett.112.120505,yoshikawa_invited_2016}. However, these experimentally generated states are Gaussian, i.e.~they can be described by a multivariate Gaussian probability distribution on the optical phase space (for formal details, see Section \ref{sec:FirstOne}). Because Gaussian statistics can easily be simulated with classical computation resources \cite{rahimi-keshari_sufficient_2016}, the use of these states in quantum computation is limited.

To reach full universal quantum computation, CV setups require at least one non-Gaussian ingredient. There have been several theoretical and experimental proposals to achieve this, ranging from ancillary Gottesman-Kitaev-Preskill (GKP) states \cite{menicucci_fault-tolerant_2014} to specific non-Gaussian gates, e.g.~\cite{arzani_polynomial_2017}. Within this paper, we focus on photon addition and subtraction as de-Gaussification techniques \cite{parigi_probing_2007}. Such techniques have, for example, proven their worth for entanglement distillation \cite{ourjoumtsev_increasing_2007,takahashi_entanglement_2010,PhysRevA.86.012328}. In particular the subtraction of a photon is in essence a simple procedure which, as originally proposed \cite{PhysRevA.55.3184}, only requires a beamsplitter and a photodetector. However, because beamsplitters are not mode-selective, this simple photon subtraction scheme will increase the impurity of the state. To avoid such incoherent mixing of modes, theory for coherent mode-selective photon subtraction was recently developed \cite{PhysRevA.89.063808,averchenko_multimode_2016}. Considerable steps have already been undertaken to implement this coherent mode-dependent photon subtraction in a quantum frequency comb \cite{ra_tomography_2017}.

In this paper, we extend the theoretical framework for such multimode photon-added and -subtracted states. Our central achievement is the derivation of the general Wigner function \cite{braunstein_quantum_2005,RevModPhys.84.621} for these states. %The Wigner function is a quasi-probability distribution \cite{braunstein_quantum_2005,RevModPhys.84.621} which is commonly used to describe CV states due to its potential for experimental quantum state tomography \cite{lvovsky_continuous-variable_2009}. 
Even though the Wigner functions for (multi)photon-added and -subtracted states are known in several specific setups, e.g.~\cite{PhysRevA.43.492, PhysRevA.55.3184, PhysRevA.71.043805, biswas_nonclassicality_2007,0953-4075-41-13-133001}, general {\em multimode} results were still lacking, even in the case of single-photon addition and subtraction.

We approach this problem from a statistical mechanics perspective, by deriving truncated correlation functions \cite{robinson_1965,petz_invitation_1990,bratteli_operator_1997,verbeure_many-body_2011} for these particular case of mode-selective photon addition and subtraction from multimode Gaussian states. Truncated correlations as such are useful witnesses for the Gaussianity of states, but they are also connected to phase-space representations. Specifically, we employ the truncated correlations to derive the characteristic function, which upon Fourier transformation gives us the Wigner function --this key result is shown in (\ref{eq:WignerFinalElegant}). In the remainder of the work, we investigate the negativity of this Wigner function, which is an important indicator of the non-classicality of the state \cite{hudson_when_1974,soto_when_1983, mandilara_extending_2009,kenfack_negativity_2004,mari_positive_2012} from a quantum probability theory perspective. %We stress that this is not necessarily equivalent to not being able to reproduce the state using classical light, which would rather be reflected in negativity or singularities in the P-function. 
Finally, we also investigate the entanglement properties that can be deduced from the Wigner function, which are ultimately the features that we want to exploit in future application in quantum technologies. This work elaborates on the details behind \cite{walschaers_entanglement_2017} and generalises the results to non-pure photon addition and subtraction.
 
The paper is structured in three major parts. In Section \ref{sec:FirstOne}, we introduce the mathematical formalism and concepts which mix techniques from quantum statistical mechanics \cite{petz_invitation_1990,bratteli_operator_1997,verbeure_many-body_2011} and quantum optics \cite{braunstein_quantum_2005,RevModPhys.84.621}. These techniques are applied in Section \ref{sec:subadd} to investigate multimode mode-selective photon addition and subtraction. To make our abstract results more concrete, we finally study two examples in Section \ref{sec:examples}: the subtraction and addition from the two-mode symmetrically squeezed vacuum, and from an experimentally obtained state. The latter is an extension of the results of \cite{walschaers_entanglement_2017}.

\section{Multimode quantum optics on phase space}\label{sec:FirstOne}

\subsection{Optical phase space and quadrature operators}

The study of continuous variable multimode quantum optics is in essence a study of quantum physics in a high-dimensional phase space. For optical systems, the relevant phase space is generated by the real and imaginary parts of the contributing electric fields, the {\em amplitude and phase quadratures}, respectively.

The modal structure of light is essential in the present contribution. A {\em mode} is simply a normalised \footnote{The modes $u_j ({\bf r}, t)$ are functions in space and time, but are typically normalised only in the spatial degrees of freedom, i.e.~$\int \abs{u_j({\bf r},t)}^2\,{\rm d}^3 {\bf r} = 1$ for any time $t$. For concrete examples and a more thorough introduction, see \cite{PhysRevA.71.013820,1367-2630-14-4-043006,PhysRevLett.100.133601}} solution $u({\bf r}, t)$ to Maxwell's equation, which has both a spatial and a temporal structure, as indicated by the arguments ${\bf r}$ and $t$, respectively. A general complex electric field $E({\bf r}, t)$ can than be represented in terms of a mode basis $\{u_j ({\bf r}, t) \mid j = 1, \dots, m\}$ as
\begin{equation}\label{eq:field}
E({\bf r}, t) = \sum_{j = 1}^m (x_j + i p_j) u_j ({\bf r}, t),
\end{equation}
where the $x_j$ and $p_j$ are the amplitude and phase quadratures, respectively. Therefore, every vector $f = (x_1, \dots, x_m, p_1, \dots, p_m)^t \in \mathbb{R}^{2m}$ can be associated with a set of phase and amplitude quadratures in the specific mode basis. The vectors space $\mathbb{R}^{2m}$ which is generated in this way is the optical phase space. As such, any vector $f \in \mathbb{R}^{2m}$ represents a classical electromagnetic field. When, in addition, $f$ is {\em normalised}, this classical electric field is associated with a new mode. Thus, it is useful to introduce \begin{equation}
{\cal N}(\mathbb{R}^{2m}) = \{f \in \mathbb{R}^{2m} \mid \norm{f} =1 \},
\end{equation}
to describe modes. However, the dimension of ${\cal N}(\mathbb{R}^{2m})$ is larger than the number of modes $m$. This is a consequence of the complex amplitude of the field, which associates two quadratures to every mode. To faithfully reproduce the properties of these complex amplitudes in (\ref{eq:field}), the phase and amplitude quadratures are connected through a symplectic structure represented by a matrix $J$ which acts on the optical phase space $\mathbb{R}^{2m}$, and has the properties:
\begin{align}
&J^2 = - \mathbb{1}, \label{eq:unitrelationJ}\\
&(f_1, J\!f_2) = - (f_2, J\!f_1)\quad \text{for all }\, f_1,f_2 \in \mathbb{R}^{2m},\label{eq:antisymmJ}
\end{align}
where $(.,.)$ denotes the standard inner product on $\mathbb{R}^{2m}$. This structure implies that the optical phase space is a phase space as also studied in analytical mechanics. An important consequence of (\ref{eq:antisymmJ}) is that $(f,J\!f) = 0$ for every $f \in \mathbb{R}^{2m}$. The orthogonal vectors $f, J\!f \in {\cal N}(\mathbb{R}^{2m})$ are associated with the same mode $u_f({\bf r}, t)$, such that the space generated by $f$ and $J\!f$ is the two-dimensional phase space which describes all possible electromagnetic fields in mode $u_f({\bf r}, t)$.

One can always construct an orthonormal {\em symplectic basis} $\mathcal{E}_s = \{e^{(1)}, \dots, e^{(m)},J\!e^{(1)},\dots, J\!e^{(m)}\}$ of the optical phase space, where $e^{(i)}$ is the basis vector which generates the phase space axis which denotes the amplitude quadrature of mode $u_i({\bf r}, t)$, whereas $J\!e^{(i)}$ generates the associated phase quadrature. The symplectic basis $\mathcal{E}_s$ of the optical phase space is directly associated with a mode basis $\{u_j ({\bf r}, t) \mid j = 1, \dots, m\}$. Hence, a change of basis in the optical phase space implies a change in mode basis.\\

% Throughout the paper, we often refer to {\em the mode $f \in {\cal N}(\mathbb{R}^{2m})$} as a shorthand for {\em the mode associated with the phase space generated by $f$ and its symplectic partner $J\!f$.}\\

When we combine the above optical phase space with the framework of statistical mechanics, we can describe classical optics setups. However, to treat problems in multimode quantum optics, we must go through the procedure of canonical quantisation. To do so, we associate a quadrature operator $Q(f)$ to each $f\in {\cal N}(\mathbb{R}^{2m})$. These operators fulfil the crucial mathematical property \begin{equation}\begin{split}\label{eq:linearityQ}
&Q(af_1+bf_2) = aQ(f_1)+bQ(f_2),
\end{split}
\end{equation} 
for all $f_1,f_2 \in {\cal N}(\mathbb{R}^{2m})$ and $a,b \in \mathbb{R}$ with $a^2+ b^2 =1$.
This property implies that the operator $Q(f)$ is independent of the basis chosen to express $f$. Moreover, these operators are governed by the canonical commutation relation \cite{Neumann_1931, petz_invitation_1990}:
\begin{equation}\label{eq:ccr}
[Q(f_1),Q(f_2)] = -2i(f_1,J\!f_2) \quad \text{for all }\, f_1,f_2 \in {\cal N}(\mathbb{R}^{2m}).
\end{equation}
We have defined (\ref{eq:ccr}) such that the operator $Q(f)$ corresponds to a quadrature operator with the shot noise equal to one. 
%\footnote{Note that the sign convention in (\ref{eq:ccr}) is different from most of the quantum optics literature. This implies that the quadrature operator associated with the phase quadrature is actually given by $-Q(J\!f)$. This also becomes apparent in the definition of the creation and annihilation operators (\ref{eq:creationAnni}).}. 

The linearity condition (\ref{eq:linearityQ}) can be extended to all $a,b \in \mathbb{R}$ to define operators $Q(\alpha)$ for non-normalised $\alpha \in \mathbb{R}^{2m}.$ This generalisation does not lead to any mathematical problems and (\ref{eq:ccr}) still holds. Physically, such different norms of $\alpha$ can be associated with rescaled quadrature measurements. In this article, the generalised quadratures will be used to limit notational overhead in the definition of the displacement operator (\ref{eq:Disp}).

\subsection{Representing quantum states}

Because quantum physics is a statistical theory, we require a mathematical object to describe the statistics of measurements: the quantum state. We focus on systems which can accurately be represented in a Hilbert space $\mathcal{H}$, on which $Q(f)$ is an (unbounded) operator. The quantum state can then be represented by a density operator $\rho$, which is positive and has $\tr \rho = 1$ \footnote{We may, in principle, also apply the methods in the present contribution to more general states and their representation of the C*-algebra of the canonical commutation relations. This allows to treat systems of infinitely many modes.}.

However, the density operator is not the most practical tool to characterise a state of a continuous variable system. Quasi-probability distributions on phase space are a common and practical alternative, not only due to their importance to interpret the fundamental physics of the state, but also because they can be measured experimentally. Throughout this paper, we will particularly emphasise the Wigner function as an important tool, because --at least for small mode numbers-- it can be experimentally reconstructed through tomographic methods.

From the mathematical point of view, we start by constructing the characteristic function in order to derive the Wigner function. To do so, we first define the displacement operator 
\begin{equation}\label{eq:Disp}D(\alpha) \equiv \exp(- i Q(J\!\alpha)/2), \quad \alpha \in \mathbb{R}^{2m}.\end{equation} 
Importantly, $\alpha$ is generally not normalised as its norm dictates the distance of the displacement. Indeed, this operator's action on a quadrature operator is given by
\begin{equation}
D(-\alpha)Q(f)D(\alpha) = Q(f) + (f,\alpha),
\end{equation}
such that the strength and direction of the displacement are given by $\norm{\alpha}$ and $\alpha/\norm{\alpha},$ respectively.
Via the displacement operator, we can introduce characteristic function as
\begin{equation}\label{eq:Chi}
\chi(\alpha) \equiv \tr \big\{ \rho D(2J\!\alpha) \big\} = \tr \big\{ \rho \exp(i Q(\alpha))\big\}.
\end{equation}
Subsequently, one can construct the Wigner function by a multi-dimensional Fourier transformation
\begin{equation}\label{eq:wignerGeneric}
W(\beta) = \frac{1}{(2\pi)^{2m}}\int_{\mathbb{R}^{2m}}{\rm d}\alpha\, \chi(\alpha)e^{- i (\alpha,\beta)}, \quad \text{for $\beta \in \mathbb{R}^{2m}$},
\end{equation}
where $\beta$ indicates a point in phase space, hence it needs not be normalised. The Wigner function shares the normalisation properties of a probability distribution and its marginals are probability measures. However, the full Wigner function is merely a quasi-probability distribution in the sense that it can assume negative values for some regions of phase space. It is this {\em negativity} which sets quantum mechanics apart from classical probabilistic theories on phase space. As such, negativity can be seen as a genuine sign of ``quantumness'', which was also associated with quantum supremacy \cite{mari_positive_2012, rahimi-keshari_sufficient_2016}.\\

Furthermore, the characteristic function can be directly linked to the {\em cumulants} of a specific quadrature measurement statistics. Indeed, $\log \chi(\alpha)$ is also known as the cumulant-generating function, which implies that
\begin{equation}\label{eq:cumulDef}
\frac{\partial^n}{\partial \lambda^n} \log \chi(\lambda f)\Big\rvert_{\lambda=0} \equiv \<Q(f)^n\>_T,
\end{equation}
with $\lambda \in \mathbb{R}$ and $f \in {\cal N}(\mathbb{R}^{2m})$. $\<Q(f)^n\>_T$ denotes the $n$th cumulant for the measurement of the quadrature $Q(f)$. A straightforward calculation \cite{petz_invitation_1990,verbeure_many-body_2011} now shows that one can recast the characteristic function in the form
\begin{equation}\label{eq:CumulantCharacteristic}\begin{split}
&\chi(\alpha) = \exp\Big\{ \sum_{n=1}^{\infty} \frac{i^n}{n!} \<Q(\alpha)^n\>_{T}\Big\}, \quad \alpha\in \mathbb{R}^{2m}.
\end{split}
\end{equation}
Thus, knowledge of all the cumulants for all the different quadratures, i.e.~for all $f \in {\cal N}(\mathbb{R}^{2m})$ and all orders $n$, implies full knowledge of the quantum state. 

We now introduce these cumulants in a more explicit form, by treating them as a special case of {\em truncated correlation functions}.

\subsection{Truncated correlation functions}\label{sec:Trunc}

The cumulants of $Q(f)$ (\ref{eq:cumulDef}) are related to the statistics of a single mode $f \in {\cal N}(\mathbb{R}^{2m})$ and do not explicitly elucidate how different modes are correlated. However, to study such questions the cumulant can be generalised to a multimode form that is commonly referred to as the {\em truncated correlation} \cite{robinson_1965} between different quadratures \footnote{In statistics literature one may also encounter the terminology ``joint cumulant', whereas in quantum field theory one also refers to the ``connected part of the correlation''.}.   

Truncated correlation functions are the multivariate extensions of cumulants and describe how quadratures for different modes are correlated. They, too, are generated using displacement operators $D(\alpha)$ (\ref{eq:Disp}). In general, we obtain
\begin{align}\label{eq:truncDef}
&\<Q(f_1) \dots Q(f_n)\>_T\\
 &\quad\equiv \frac{\partial \log \tr(\rho D(- 2 \lambda_1 J\!f_1) \dots D(-2 \lambda_n J\!f_n))}{\partial \lambda_1 \dots \partial \lambda_n}\Big\rvert_{\lambda_1 = \dots = \lambda_n =0},\nonumber
\end{align}
which can be related to $\chi(\alpha)$ in (\ref{eq:Chi}) through the identity $D(\alpha)D(\beta)=D(\alpha + \beta) \exp\{- i (\alpha, J\beta)/4\}$. 

In an experimental setting, it is more practical to obtain the truncated correlation functions by jointly measuring distinct quadratures and following a recursive recipe:
\begin{align}
&\<Q(f_1)\>_T = \tr \{\rho Q(f_1) \} \label{eq:TruncationScheme}\\
&\<Q(f_1)Q(f_2)\>_T = \tr \{\rho Q(f_1)Q(f_2) \} - \<Q(f_1)\>_T\<Q(f_2)\>_T\nonumber \\
&\<Q(f_1)Q(f_2)Q(f_3)\>_T =  \tr \{\rho Q(f_1)Q(f_2)Q(f_3) \} \nonumber\\
&\qquad\qquad\qquad\qquad\qquad - \<Q(f_1)Q(f_2)\>_T\<Q(f_3)\>_T\nonumber\\
&\qquad\qquad\qquad\qquad\qquad-\<Q(f_1)Q(f_3)\>_T\<Q(f_2)\>_T \nonumber\\
&\qquad\qquad\qquad\qquad\qquad-\<Q(f_2)Q(f_3)\>_T\<Q(f_1)\>_T\nonumber\\
&\qquad\qquad\qquad\qquad\qquad-\<Q(f_1)\>_T\<Q(f_3)\>_T\<Q(f_2)\>_T  \nonumber \\
&\qquad\qquad\qquad\dots \text{\em et cetera.}\nonumber
\end{align}
These truncated correlation functions are experimentally measurable through, for example, multimode homodyne measurement \cite{PhysRevLett.84.5748,armstrong_programmable_2012,cai_reconfigurable_2016}. By expanding $\alpha \in \mathbb{R}^{2m}$ in a specific mode basis in (\ref{eq:CumulantCharacteristic}), the role of truncated correlations becomes apparent. Thus, the set of truncated correlations forms an important tool for characterisation. More specifically, one may wonder to what order one needs to measure these correlations to extract a given property of the state. Such a property, which is of special interest throughout this text, is the state's Gaussianity.

\subsection{Gaussian states}\label{sec:GaussianStates}

Of particular importance in quantum optics are the Gaussian states. In the broad sense, a state of a CV system is said to be Gaussian if it induces Gaussian statistics in all modes. This implies that the function $\chi$, and hence also the Wigner function, is a multivariate Gaussian \cite{braunstein_quantum_2005,RevModPhys.84.621}:
\begin{align}\label{eq:GaussianChar}
\chi_G(\alpha) &= \exp\Big\{-\frac{(\alpha, V \alpha)}{2} + i (\xi,\alpha)\Big\},\\
W_G(\beta) &= \frac{\exp\left\{-\frac{1}{2}\Big((\beta-\xi), V^{-1} (\beta-\xi)\Big)\right\}}{(2\pi)^m\sqrt{\det V}} \label{eq:GaussianWig}
\end{align}
where $\xi$ is a vector which describes the states displacement, and $V$ is referred to as the {\em covariance matrix}. Therefore $V$ is positive semi-definite matrix on $\mathbb{R}^{2m}$, which describes the correlations between different field quadratures in a specific mode basis. A crucial demand for this $V$ to be associated with a well-defined quantum state is given by \cite{petz_invitation_1990}
\begin{equation}\label{eq:Heisenberg}
(f_1, V f_1)(f_2, V f_2) \geqslant \abs{(f_1, J\!f_2)}^2, \quad \text{for all }\, f_1,f_2 \in {\cal N}(\mathbb{R}^{2m}),
\end{equation}
which is the multimode version of Heisenberg's uncertainty relation. Alternatively, the properties of $V$ can also be expressed by the condition $V + iJ \geqslant 0$.

The insertion of (\ref{eq:GaussianChar}) in (\ref{eq:truncDef}) imposes important conditions upon the truncated correlations of Gaussian states. At first it is directly obtained that
\begin{align}
&\<Q(f)\>_T = (\xi,f),\\
&\<Q(f_1)Q(f_2)\>_T = (f_1,V f_2) - i (f_1,J\!f_2).\label{eq:CovMat1}
\end{align}
Furthermore, we deduce the general condition that {\em for a Gaussian state}
\begin{equation}\label{eq:GaussianCor}
\<Q(f_1)\dots Q(f_n)\>_T = 0, \quad n>2,
\end{equation}
for all $f_1, \dots, f_n \in {\cal N}(\mathbb{R}^{2m})$. The implication of (\ref{eq:GaussianCor}) is that all non-Gaussian states must have non-zero truncated correlations. Furthermore, it was shown \cite{robinson_theorem_nodate} that for non-Gaussian states, there is never an order from which onward the truncated correlation functions become zero. Therefore these functions are ideal tools for the operational characterisation of non-Gaussian states. Specifically in multimode systems where full tomographies tend to be completely unfeasible, they are an experimentally accessible alternative.

\subsection{Entanglement}\label{sec:Entanglement}

In the context of quantum physics, one often associates correlations to the study of entanglement which is commonly seen as an important resource for quantum computation and quantum communication. The profound advantage of CV quantum optics, is the simplicity with which Gaussian entanglement between modes can be generated. CV entanglement is strongly dependent of the mode basis in which the problem is described. To see this, it is instructive to consider an arbitrary symplectic basis ${\cal B} = \{b^{(1)}, \dots, b^{(m)}, J\!b^{(1)},\dots, J\!b^{(m)}\}$ of $\mathbb{R}^{2m}$ and express $\beta \in \mathbb{R}^{2m}$ in (\ref{eq:wignerGeneric}) in this basis:
\begin{equation}
\beta = \sum_{i=1}^m \zeta^{(i)}_x b^{(i)} + \zeta^{(i)}_pJ\!b^{(i)}.
\end{equation}
In this mode basis, the Wigner function is a function of the $\zeta$-variables, i.e.~$W(\beta) = W(\zeta^{(1)}_x, \dots, \zeta^{(m)}_x, \zeta^{(1)}_p, \dots, \zeta^{(m)}_p)$ \footnote{A different choice of mode basis ${\cal B}$ changes the form of the Wigner function. Thus, physical properties such as entanglement, which depend on the form of the Wigner function depend on the chosen mode basis.}. 

We refer to a CV state as {\em fully separable} in the mode basis ${\cal B}$ when its Wigner function can be written as
\begin{equation}\begin{split}\label{eq:entGen}
 &W(\zeta^{(1)}_x, \dots, \zeta^{(m)}_x, \zeta^{(1)}_p, \dots, \zeta^{(m)}_p)\\
 &\qquad\qquad\qquad = \int {\rm d}\lambda \, p(\lambda) \prod_{i=1}^m W_{\lambda}(\zeta^{(i)}_x,\zeta^{(i)}_p),
 \end{split}
 \end{equation}
 a statistical mixture of a product of single-mode Wigner functions. To obtain a statistical mixture, $\lambda$ must correspond to a way of labelling states, and $p(\lambda)$ is a probability distribution on this set of labels. Any state for which (\ref{eq:entGen}) does not hold is said to be {\em entangled} in mode basis ${\cal B}$. Note that one can introduce more refined terminology for multimode entanglement \cite{PhysRevLett.114.050501}. Analysing such different types of multipartite entanglement, however, falls beyond the scope of this work.
 
 It is natural to ask whether there always exists a mode basis in which the state is separable. We will provide a negative answer to this question by showing that this is generally not the case for photon-added and subtracted states. If we can construct a mode basis for which (\ref{eq:entGen}) holds, we will refer to the state as {\em passively separable}, to highlight that any entanglement present in a specific mode basis can be undone by passive linear optics. States which are not passively separable are now referred to as {\em inherently entangled}. \\

We show now that the Gaussian states of Section \ref{sec:GaussianStates} are always passively separable, by using the properties of their covariance matrices. The Wigner function (\ref{eq:GaussianWig}) of a non-displaced (i.e.~$\xi = 0$) Gaussian state $\rho_G$ is completely governed by the positive semidefinite covariance matrix $V$, which can be decomposed as $V= S^t\Delta S$ through the Williamson decomposition. Here $S$ is a symplectic matrix and $\Delta \geqslant \mathbb{1}$ is diagonal (the diagonal elements of $\Delta$ are known as the symplectic spectrum). Because $S$ is symplectic, it can be further decomposed with the Bloch-Messiah decomposition: $S = O' K O$, where $O$ and $O'$ are orthogonal and symplectic, and $K$ is diagonal and symplectic. This now allows us to rewrite $V = O^t K V_{\rm th} K O$, where $V_{\rm th} = O'^t \Delta O' \geqslant \mathbb{1}$ is the covariance matrix of a thermal state. We use this structure to separate the covariance into a pure part and added classical noise, \begin{equation}V = V_s + V_c.\label{eq:thisdecomphere}\end{equation} Here, $V_s = O^t K^2 O$ is the covariance matrix of a pure squeezed vacuum state $\rho_s$, and $V_c = O^t K (V_{\rm th} - \mathbb{1}) K O$ is the covariance matrix of the additional noise. Note that, a priori, $V_c$ does not fulfil (\ref{eq:Heisenberg}) and is therefore not the covariance matrix of a quantum state. We can think of the state characterised by $V$ as being generated by injecting $\rho_s$ into a noisy Gaussian channel \cite{PhysRevA.69.052320,cerf_gaussian_2007}. We obtain that
\begin{equation}\label{eq:notoriousDecomp}
\rho_G = \int_{\mathbb{R}^{2m}} {\rm d}^{2m}\xi' D(\xi')\rho_s D(-\xi') \frac{\exp\left\{-\frac{(\xi, V_c^{-1 }\xi)}{2}\right\}}{(2\pi)^m \sqrt{\det V_c}},
\end{equation}
which implies that the Wigner function for $\rho_G$ can be represented by
\begin{equation}\label{eq:GaussianMix}
W_G(\beta) = \int {\rm d}^{2m}\xi \, W_s(\beta-\xi) p_c(\xi),
\end{equation}
where 
\begin{equation}
p_c(\xi) = \frac{\exp\left\{-\frac{(\xi, V_c^{-1 }\xi)}{2}\right\}}{(2\pi)^m \sqrt{\det V_c}},
\end{equation}
and 
\begin{equation}
W_s(\beta) = \frac{\exp\left\{-\frac{1}{2}\Big(\beta, V^{-1}_s \beta\Big)\right\}}{(2\pi)^m\sqrt{\det V_s}}. 
\end{equation}
The Bloch-Messiah decomposition naturally gives a specific mode basis (obtained through the orthogonal transformation $O$) in which $W_s$ factorises for any $\xi \in \mathbb{R}^{2m}$. In this mode basis, the Wigner function (\ref{eq:GaussianMix}) has the form (\ref{eq:entGen}), which implies that the state $\rho_G$ is passively separable. Thus, any entanglement that is present in the original mode basis can be undone by a passive linear optics circuit that is described by $O^t$, or by measuring quadratures in mode basis associated with $O$.

Thus we provided an explicit construction of a linear optics operation to render a given Gaussian state separable. For mixed states this is typically not the only linear optics operation that can undo entanglement. Indeed, the core ingredient of the decomposition (\ref{eq:notoriousDecomp}) is (\ref{eq:thisdecomphere}), such that for every pure state covariance matrix $V'_s \leqslant V$, we can set $V_c = V - V'_s$ in (\ref{eq:thisdecomphere}). Because $V'_s$ characterises a pure, Gaussian state, we can find a mode basis of symplectic eigenvectors of $V'_s$. We can thus simply diagonalise $V'_s = O'^t K'^2 O'$, where $O'^t$ describes an alternative linear optics circuit that can undo entanglement in the Gaussian state. The above method, using Williamson and Bloch Messiah, shows that such a $V'_s$ always exists.\\ 

For simplicity, we assumed that $\rho_G$ was non-displaced. However, the argument is straightforwardly extended to displaced Gaussian states by letting the displacement operator act on $\rho_G$.

\section{Single-photon added and subtracted Gaussian states}\label{sec:subadd}

\subsection{Induced correlations}

As we argued in the introduction of this paper, non-Gaussianity is a crucial ingredient to achieve universal quantum computation. Moreover, for technological applications, it is essential that the complexity of any quantum device can be increased, hence requiring a sense of scalability. In CV quantum optics, we first and foremost consider such scalability in the number of modes. Therefore, we must consider a multimode setup, in which we can incorporate a non-Gaussian operation. From an experimental perspective, a promising procedure to fulfil these conditions is mode-selective photon subtraction \cite{averchenko_multimode_2016,PhysRevA.89.063808,ra_tomography_2017} or addition \cite{zavatta_experimental_2007,parigi_probing_2007}. 

In this contribution, we limit ourselves to the subtraction or addition of a single photon in a setup with an arbitrary mode number $m$. To effectively model the associated subtraction and addition procedures, we must introduce the annihilation and creation operators for an arbitrary vector in phase space $g \in {\cal N}(\mathbb{R}^{2m})$, $a(g)$ and $a^{\dag}(g)$, respectively. In our framework, they are defined as
\begin{equation}\label{eq:creationAnni}\begin{split}
a^{\dag}(g) &\equiv \frac{1}{2}\big(Q(g) - i Q(J\!g)\big), \\
 a(g) &\equiv \frac{1}{2}\big(Q(g) + i Q(J\!g)\big),
 \end{split}
\end{equation}
from which we directly obtain an alternative version of the canonical commutation relation (\ref{eq:ccr}),
\begin{equation}
[a(g_1), a^{\dag}(g_2)] = (g_1,g_2) + i (g_1, J\!g_2).
\end{equation}
These operators create or annihilate photons in a specific mode, represented by $g$. However, $g$ is a vector in the $2m$ dimensional phase space, whereas there are only $m$ modes. Therefore, we stress that $a^{\dag}(J\!g) = -i\, a^{\dag}(g)$, such that the photons created by the operators $a^{\dag}(g)$ and $a^{\dag}(J\!g)$ clearly only differ by a global phase. As global phases have no physical importance in quantum physics, the creation operators $a^{\dag}(g)$ and $a^{\dag}(J\!g)$ really create a photon in the same mode. 

We now focus on an arbitrary non-displaced Gaussian state, which we formally describe by a density matrix $\rho_G$, and convert it to a non-Gaussian state by means of mode-selective photon-addition or -subtraction in a mode $g \in {\cal N}(\mathbb{R}^{2m})$. As was argued in Section \ref{sec:GaussianStates}, this non-displaced Gaussian state can be completely characterised by its covariance matrix $V$. The new, photon-added and -subtracted states' density operators are then given by
\begin{equation}\label{eq:photonSubRho}
\rho_- = \frac{a(g)\rho_G a^{\dag}(g)}{\<\hat{n}(g)\>_G},
\end{equation}
for subtraction, and 
\begin{equation}\label{eq:photonAddRho}
\rho_+ = \frac{a^{\dag}(g)\rho_G a(g)}{\<\hat{n}(g)\>_G+1},
\end{equation}
for addition. We introduced the notation $\<.\>_G \equiv \tr(\rho_G .)$ for the expectation values in the state $\rho_G$, and $\hat{n}(g)\equiv a^{\dag}(g)a(g)$ for the number operator that counts the number of photons in the mode $g \in {\cal N}(\mathbb{R}^{2m})$. \\

In order to characterise the non-Gaussian features of the system, we follow the ideas of Section \ref{sec:Trunc} and evaluate the truncated correlation functions. If the state is, indeed, non-Gaussian, we should obtain non-zero values for the truncated correlation functions of some order beyond than two. However, because we intend to use the recursive procedure of Section \ref{sec:Trunc}  to evaluate the correlations, it is instructive to start by evaluating the two-point correlation $\<Q(f_1)Q(f_2)\>_T$, for arbitrary $f_1, f_2 \in {\cal N}(\mathbb{R}^{2m})$. Because the state is non-displaced, by definition $\<Q(f)\>_T = 0,$ and we obtain that for the photon-subtracted (hence the superscript ``$-$'') state
\begin{align}
\<Q(f_1)Q(f_2)\>^{-}_T &= \tr\{\rho_{-} Q(f_1)Q(f_2)\} \\&= \frac{\<a^{\dag}(g)Q(f_1)Q(f_2)a(g)\>_G}{\<\hat{n}(g)\>_G},
\end{align}
where we used (\ref{eq:photonSubRho}) and the cyclic property of the trace. Analogously, for photon-addition we obtain
\begin{align}
\<Q(f_1)Q(f_2)\>^{+}_T = \frac{\<a(g)Q(f_1)Q(f_2)a^{\dag}(g)\>_G}{\<\hat{n}(g)\>_G+1},
\end{align}
The property (\ref{eq:GaussianCor}) for non-displaced Gaussian states implies that expectation values of products of quadrature operators factorises in pairs \cite{PhysRev.130.2529,robinson_1965,petz_invitation_1990,verbeure_many-body_2011}. Combining this with the definition (\ref{eq:creationAnni}) for the creation and annihilation operators in terms of quadratures, and with the linearity of the trace, we find
\begin{align}\label{eq:TwoPointCorSubt}
\<Q(f_1)Q(f_2)\>^{-}_T = &\frac{\<a^{\dag}(g)Q(f_1)\>_G\<Q(f_2)a(g)\>_G}{\<\hat{n}(g)\>_G}\nonumber\\ &+ \frac{\<a^{\dag}(g)Q(f_2)\>_G\<Q(f_1)a(g)\>_G}{\<\hat{n}(g)\>_G}\\ &+ \<Q(f_1)Q(f_2)\>_G.\nonumber
\end{align}
and
\begin{align}\label{eq:TwoPointCorAdd}
\<Q(f_1)Q(f_2)\>^{+}_T = &\frac{\<a(g)Q(f_1)\>_G\<Q(f_2)a^{\dag}(g)\>_G}{\<\hat{n}(g)\>_G+1}\nonumber\\ &+ \frac{\<a(g)Q(f_2)\>_G\<Q(f_1)a^{\dag}(g)\>_G}{\<\hat{n}(g)\>_G+1}\\ &+ \<Q(f_1)Q(f_2)\>_G.\nonumber
\end{align}
To proceed in the evaluation, we use (\ref{eq:CovMat1}) and (\ref{eq:creationAnni}) to obtain
\begin{align}
\<a^{\dag}(g) Q(f)\>_G &= \frac{1}{2}\Big((f,[V-\mathbb{1}]g) - i (f,[V-\mathbb{1}]Jg)\Big),\nonumber\\
\<Q(f)a^{\dag}(g)\>_G &= \frac{1}{2}\Big( (f, [V+\mathbb{1}]g) - i (f,[V+\mathbb{1}]Jg)\Big),\nonumber\\
\<Q(f)a(g)\>_G &= \frac{1}{2}\Big( (f,[V-\mathbb{1}]g) + i (f,[V-\mathbb{1}]Jg)\Big),\nonumber\\
\<a(g) Q(f)\>_G &= \frac{1}{2}\Big((f, [V+\mathbb{1}] g) + i \big[(f,[V+\mathbb{1}]Jg)\big]\Big),\nonumber\\
\<\hat{n}(g)\>_G &= \frac{1}{4} \Big((g,V g) + (Jg,V Jg) - 2\Big).\label{eq:theseguys}
\end{align}
When we insert these results in (\ref{eq:TwoPointCorSubt}), we ultimately obtain that for the photon-subtracted non-displaced Gaussian state\begin{align}
\label{eq:BoomAg}
&\<Q(f_1)Q(f_2)\>^{\pm}_T = \<Q(f_1)Q(f_2)\>_G + (f_1,A^{\pm}_g f_2),\\
&\text{with}\quad (f_1,A^{-}_gf_2) \equiv \frac{\<a^{\dag}(g)Q(f_1)\>_G\<Q(f_2)a(g)\>_G}{\<\hat{n}(g)\>_G}\nonumber\\ &\qquad \qquad \qquad \qquad+\frac{\<a^{\dag}(g)Q(f_2)\>_G\<Q(f_1)a(g)\>_G}{\<\hat{n}(g)\>_G},\nonumber\\
&\text{and}\quad (f_1,A^{+}_gf_2) \equiv \frac{\<a(g)Q(f_1)\>_G\<Q(f_2)a^{\dag}(g)\>_G}{\<\hat{n}(g)\>_G+1}\nonumber\\ &\qquad \qquad \qquad \qquad+\frac{\<a(g)Q(f_2)\>_G\<Q(f_1)a^{\dag}(g)\>_G}{\<\hat{n}(g)\>_G+1},\nonumber
\end{align}
$A^{\pm}_g$ is a matrix which acts on the space $\mathbb{R}^{2m}$. Inserting (\ref{eq:theseguys}) in (\ref{eq:BoomAg}) directly leads to
\begin{equation}\label{eq:AgMat}A^{\pm}_g = 2\frac{(V \pm \mathbb{1})(P_g + P_{Jg})(V \pm \mathbb{1})}{\tr\{ (V \pm \mathbb{1})(P_g + P_{Jg})\}}.
\end{equation}
Here we introduced $P_g$ and $P_{Jg}$ as the projectors on the vectors $g$ and $Jg$, respectively. This implies that $P_g + P_{Jg}$ is the projector on the two-dimensional phase space, associated with the mode in which the photon was subtracted. It can directly be verified that $A_g^{\pm}$ describes additional correlations between quadratures that are induced by the photon-subtraction or -addition process. Ultimately, these additional correlations are completely determined by the mode $g$ from which the photon is subtracted, and the correlation matrix $V$ of the initial non-displaced Gaussian state $\rho_G$.\\

Experimentally, however, it is hard to guarantee that (\ref{eq:photonSubRho}) and (\ref{eq:photonAddRho}) are the exact states which we obtain. In general, the subtraction process adds some degree of impurity. There are various sources of impurities in an experimental context, ranging from photon-losses to contributions of higher photon-numbers in the subtraction \cite{ra_tomography_2017}, which go beyond the scope of the present work. Nevertheless, we consider one important type of impurity in the subtraction process, related to lack of control of the mode-selectivity \cite{averchenko_multimode_2016}. In the most extreme case, one may think of photon subtraction by means of a beamsplitter, where it is impossible to infer from which mode the photon originated in the case of co-propagating modes. In general terms, it is hard to control exactly in which mode the photon is added or subtracted \cite{averchenko_multimode_2016}. This implies that we have to deal with a mixture of the form
\begin{equation}\label{eq:photonSubRhoMix}
\begin{split}
&\rho_{-} =  \frac{\sum_{k} \gamma_k a(g_k)\rho_G a^{\dag}(g_k)}{\sum_{k} \gamma_k\<\hat{n}(g_k)\>_G},\\& \quad \text{with} \, \sum_k \gamma_k =1, \, \text{ and }\, \gamma_k\geqslant 0,
\end{split}
\end{equation}
for subtraction, or 
\begin{equation}\label{eq:photonAddRhoMix}
\begin{split}
&\rho_{+} =  \frac{\sum_k \gamma_k a^{\dag}(g_k)\rho_G a(g_k)}{1+\sum_{k} \gamma_k \<\hat{n}(g_k)\>_G},\\& \quad \text{with} \, \sum_k \gamma_k =1, \, \text{ and }\, \gamma_k\geqslant 0,
\end{split}
\end{equation}
for addition. The details of the participating modes and the $\gamma_k$ depend strongly on the experimental setup and can be estimated through a detailed modelling \cite{averchenko_multimode_2016,PhysRevA.89.063808}. 

Through the linearity of the expectation value, we can directly verify that 
\begin{equation}\label{eq:BoomAmix}
\begin{split}
&\<Q(f_1)Q(f_2)\>^{\pm}_T = \<Q(f_1)Q(f_2)\>_G + (f_1, A^{\pm}_{\rm mix} f_2),
\end{split}
\end{equation}
where
\begin{equation}\label{eq:AmixMat}\begin{split}&A^{\pm}_{\rm mix} \\&\quad= 2(V \pm \mathbb{1})\frac{ \sum_k \gamma_k (P_{g_k} + P_{Jg_k})}{\tr\{ (V \pm \mathbb{1})\sum_k \gamma_k (P_{g_k} + P_{Jg_k})\}} (V \pm \mathbb{1}).\end{split}
\end{equation}
%Moreover, we can see that the structure of (\ref{eq:ThisEquationInApp}) is also obtained in the mixed case
%\begin{equation}\begin{split}\label{eq:ingredientTwo}
%\tr\{\rho& Q(f_1)\dots Q(f_{2k}) \}\\
%=&\sum_{p \in {\cal P}^{(2)}}\prod_{i \in p} \<Q(f_{i_1})Q(f_{i_2})\>_G\\
%&+ \sum_{p \in {\cal P}^{(2)}}\sum_{i \in p} \Bigg( (f_{i_1}, A^{\pm}_{\rm mix} f_{i_2})\prod_{j \in p \setminus i} \<Q(f_{j_1})Q(f_{j_2})\>_G\Bigg),\end{split}
%\end{equation}
%where $A_{\rm mix}$ can both refer to photon-subtraction or -addition. With (\ref{eq:BoomAmix}) and (\ref{eq:ingredientTwo}) we have all the necessary ingredients for the derivation of (\ref{sec:ProofCor}). Therefore, we find that also in the mixed case
We use (\ref{eq:BoomAmix}) to evaluate the higher-order truncated correlations in Appendix \ref{sec:ProofCor}. This leads to the remarkable result that these truncated correlations, too, are governed by the matrix $A^{\pm}_{\rm mix}$. As a final result, we obtain
\begin{equation}\label{eq:TruncFinalMix} \boxed{\begin{split}
\<&Q(f_1)\dots Q(f_{2k})\>_T\\
&\qquad = (-1)^{k-1} (k-1)! \sum_{p \in {\cal P}^{(2)}}\prod_{i \in p} (f_{i_1}, A^{\pm}_{\rm mix} f_{i_2}),\\
\<&Q(f_1)\dots Q(f_{2k-1})\>_T = 0,
\end{split}}
\end{equation}
for all $k > 1$. ${\cal P}^{(2)}$ indicates the set of all pair-partitions, i.e.~all the ways of dividing the set $\{f_1, \dots, f_{2k}\}$ up in $k$ pairs. In literature, e.g.~\cite{mansour_combinatorics_2013}, this partition is also known as a perfect matching. For even orders, the truncated correlations (\ref{eq:TruncFinalMix}) are generally non-zero. This is a clear statistical signature of the non-Gaussian character of the state.

\subsection{Phase space representations}\label{sec:phaseSpace}

\subsubsection{Multimode Wigner function}
To highlight that the above truncated correlation functions grant us full knowledge of the quantum state, we use them to construct the quantum characteristic function for the photon-subtracted state (\ref{eq:photonSubRho}) by virtue of (\ref{eq:CumulantCharacteristic}). To do so, we need to know the state's cumulants, which are the truncated correlations (\ref{eq:TruncFinalMix}) for $f_1 = f_2 = \dots = f_{2k} = f$. Central in this evaluation is that every pair-partition $p \in {\cal P}^{(2)}$ in (\ref{eq:TruncFinalMix}) contributes the same term, $(f,A^{\pm}_{\rm mix}f)^k$, because contribution of each mode is the same. We only need to count the number of pair-partitions to know which combinatorial factor to add. %This number is well-known in combinatorics and is given by the double factorial, in our case $(2k-1)!! = (2k-1)(2k-3)(2k-5)\dots 3$. 
We find that the cumulant is given by
\begin{align}
&\<Q(f)^{2k}\>_T %&= (-1)^{k-1} (k-1)! (2k - 1)!! (f,A^{\pm}_{\rm mix}f)^k\nonumber \\&\qquad+ (f,Vf) \delta_{k,1}\\
 =  (-1)^{k-1} \frac{(2k - 1)!}{2^{k-1}} (f,A^{\pm}_{\rm mix}f)^k\\&\qquad\qquad\qquad + (f,Vf) \delta_{k,1},\nonumber\\
&\<Q(f)^{2k-1}\>_T = 0.\label{eq:cumulZero}
\end{align}
%It is instructive and elegant to see that 
%\begin{equation}
%A^s_g(f,f)= 2\frac{(f,[V-\mathbb{1}]g)^2 + (f,[V-\mathbb{1}]Jg)^2 }{(g,Vg) + (Jg, V Jg) - 2},
%\end{equation}
%and
%\begin{equation}
%A^a_g(f,f)= 2\frac{(f,[V+\mathbb{1}]g)^2 + (f,[V+\mathbb{1}]Jg)^2 }{(g,Vg) + (Jg, V Jg) + 2},
%\end{equation} 
It now remains to evaluate the series in Eq.~(\ref{eq:CumulantCharacteristic}), for which we find that 
\begin{equation}\begin{split}\label{eq:seriesExpnasion}
 \sum_{n=1}^{\infty}& \frac{i^n\lambda^n}{n!} \<Q(f)^n\>_{T}\\
  &= -\frac{\lambda^2 (f,Vf)}{2} \\&\quad+ \sum_{k=1}^{\infty} \frac{i^{2k}\lambda^{2k}}{(2k)!}(-1)^{k-1} \frac{(2k - 1)!}{2^{k-1}} (f,A^{\pm}_{\rm mix}f)^k\\
 &=-\frac{\lambda^2 (f,Vf)}{2} - \sum_{k=1}^{\infty} \frac{1}{k}\Bigg(\frac{\lambda^2 (f,A^{\pm}_{\rm mix}f)}{2}\Bigg)^k,
 \end{split}
 \end{equation}
 where the series was already rewritten to only sum over the even cumulants since all odd contributions are zero.The final series in (\ref{eq:seriesExpnasion}) is subtle because it does not necessarily converge. However, $\chi(\lambda f)$ maps the points of phase space to the complex plane, as such we may resort to an analytical continuation of the series to obtain that 
\begin{align}
\chi(\lambda f) &= \exp\Bigg\{-\frac{\lambda^2}{2} (f,Vf) - \sum_{k=1}^{\infty} \frac{1}{k}\Bigg(\frac{\lambda^2 (f,A^{\pm}_{\rm mix}f)}{2}\Bigg)^k\Bigg\} \nonumber\\
&= \Bigg(1 - \frac{\lambda^2 (f,A^{\pm}_{\rm mix}f)}{2} \Bigg)\exp\Bigg\{-\frac{\lambda^2}{2} (f,Vf)\Bigg\}.
\end{align}
Therefore we have obtained the quantum characteristic function (\ref{eq:Chi}) to fully characterise the states. In principle, this also allows us to derive the Wigner function by means of a multi-dimensional Fourier transformation. We may formally write the Wigner function as
\begin{equation}\begin{split}\label{eq:wignerGen}
W^{\pm}(\beta) &= \frac{1}{(2\pi)^{2m}}\int_{\mathbb{R}^{2m}}{\rm d}\alpha\, \chi(\alpha)e^{- i (\alpha,\beta)}\\
&= \frac{1}{(2\pi)^{2m}} \int_{\mathbb{R}^{2m}}{\rm d}\alpha \Bigg(1 - \frac{(\alpha,A^{\pm}_{\rm mix}\alpha)}{2} \Bigg)\\
&\qquad\qquad\qquad\times\exp\Bigg\{-\frac{(\alpha,V\alpha)}{2} - i (\alpha,\beta)\Bigg\},
\end{split}
\end{equation}
This Fourier transform is explicitly computed in Appendix \ref{app:B}, and leads to
\begin{equation}\label{eq:WignerFinalElegant}
\boxed{W^{\pm}(\beta) = \frac{1}{ 2} \Big((\beta, V^{-1} A^{\pm}_{\rm mix} V^{-1} \beta) -  \tr(V^{-1}A^{\pm}_{\rm mix}) + 2 \Big)W_G(\beta),}
\end{equation}
where $W_G(\beta)$ is the Wigner function of the initial Gaussian state (\ref{eq:GaussianWig}) before the addition or subtraction of the photon. This now gives us the full, multimode, Wigner functions of a non-displaced photon-added or -subtracted state. We observe that the general structure of the Wigner function is given by a multivariate polynomial of order two, multiplied by the Gaussian Wigner function of the initial state.

\subsubsection{Negativity}

The negativity of the Wigner function is often seen as a genuine quantum feature in CV systems. With (\ref{eq:WignerFinalElegant}) we have all the tools at hand to analyse such features in the Wigner function. 

At first, we note that the Wigner function (\ref{eq:WignerFinalElegant}) is negative if and only if there are vectors $\beta \in \mathbb{R}^{2m}$ for which
\begin{equation}\label{eq:NegativityCondition}
(\beta, V^{-1} A^{\pm}_{\rm mix} V^{-1} \beta) -  \tr(V^{-1}A^{\pm}_{\rm mix}) + 2 \leqslant 0.
\end{equation}
However, it is directly verified that $V^{-1} A^{\pm}_{\rm mix} V^{-1}$ is a positive-semidefinite matrix, hence 
\begin{equation}
(\beta, V^{-1} A^{\pm}_{\rm mix} V^{-1} \beta) \geqslant 0, \quad \text{ for all $\beta \in \mathbb{R}^{2m}$.}
\end{equation}
Therefore, the {\em necessary and sufficient} condition for the existence of negative values of the Wigner function is
\begin{equation}\label{eq:conditionNegat}
\boxed{
 \tr(V^{-1}A^{\pm}_{\rm mix}) \geqslant 2.}
\end{equation}
By setting $\beta = 0$ in (\ref{eq:NegativityCondition}) we clearly see that (\ref{eq:conditionNegat}) is, indeed, a sufficient condition. Through (\ref{eq:AmixMat}), we can rephrase this condition as
\begin{equation}\label{eq:thiscoolone}
\begin{split}
&\sum_{k} \gamma_k \big[(g_k,V^{-1} g_k) + (J\!g_k,V^{-1} J\!g_k)\big] > 2, \quad\text{for subtraction,}\\
&\sum_{k} \gamma_k \big[(g_k,V^{-1} g_k) + (J\!g_k,V^{-1} J\!g_k)\big] > -2, \quad\text{for addition,}
\end{split}
\end{equation}
which is automatically fulfilled for photon addition. Hence, we formally show that photon addition to a non-displaced Gaussian state {\em always} induces a negative Wigner function, even when the initial state and the addition process are mixed. On the other hand, for photon subtraction the condition for negativity of the Wigner function can be violated when there is too much thermal noise compared to the amount of squeezing (see, e.g., the example in Section \ref{sec:ex1}).

%With (\ref{eq:AMat}), we can recast the condition in the form 
%making it apparent that this condition which is trivially fulfilled for the case of photon-addition (``$+$''). Hence, independent of the initial Gaussian state, the addition of a photon will always lead to a negative Wigner function. On the other hand, given an initial covariance matrix $V$, (\ref{eq:thiscoolone}) can be used to determine from which modes $g \in \mathbb{R}^{2m}$ a photon can be subtracted in order to render the Wigner function negative.\\

Finally, we emphasise that the equation
\begin{equation}\label{eq:ZeroRegion}
(\beta, V^{-1} A^{\pm}_{\rm mix} V^{-1} \beta) =  \tr(V^{-1}A^{\pm}_{\rm mix}) - 2
\end{equation}
defines the manifold of zeros of the Wigner function. Specifically equation (\ref{eq:ZeroRegion}) generates a multidimensional ellipsoid. The details of the manifold depend strongly on the details of the subtraction or addition process, and on the covariance matrix $V$. However, as expected, the general condition for equation (\ref{eq:ZeroRegion}) to have solutions is also given by (\ref{eq:conditionNegat}).

%
%In the case where $V$ can be diagonalised by a set of symplectic eigenvectors ${\cal E}_s = \{e^{(1)}_x, \dots e_x^{(m)},e_p^{(1)},\dots e^{(m)}_p\}$, it is convenient to rewrite
%\begin{equation}\label{eq:NegativeWigner2}
% \tr(V^{-1}A_{\rm mix})=\sum_{q \in \{x,p\}}\sum_{j=1}^m \frac{(e^{(j)}_q, e^{(j)}_q)}{v^{(j)}_q}.
%\end{equation}
%Eq.~(\ref{eq:NegativeWigner2}) only requires us to know that symplectic eigenvectors and eigenvalues of the covariance matrix $V$ that determines the initial Gaussian state. In this supermode basis, we find $A_g(e^{(j)}_q, e^{(j)}_q) = \<Q(e^{(j)}_q)^2\>_T - v^{(j)}_q$. Therefore, once the supermodes are known, all that remains is to measure the variance of the respective quadratures after photon-subtraction or -addition. Then, one can use (\ref{eq:conditionNegat}) to verify that the states Wigner function has a negative region.
% 

\subsubsection{Entanglement}\label{sec:EntanglementAddSub}

In this section we elaborate on the passive separability of the Wigner function (\ref{eq:WignerFinalElegant}). First, we prove that, whenever a photon is added or subtracted to or from a mode which is not entangled to any other modes in the initial Gaussian state, the resulting photon-added or -subtracted state will remain passively separable. We then prove for pure states, that subtraction or addition of a photon in any other mode renders the state inherently entangled.\\

For any possible decomposition (\ref{eq:thisdecomphere}) of the Gaussian state's covariance matrix $V$, we may use (\ref{eq:notoriousDecomp}) to write the photon-subtracted state as
\begin{equation}
\begin{split}\label{eq:rhoExpandedSub}
\rho^- &= \frac{a(g)\rho_G a^{\dag}(g)}{\<\hat{n}(g)\>_G} \\ &= \frac{1}{\<\hat{n}(g)\>_G} \int {\rm d}^{2m}\xi \, a(g)D(\xi)\rho_sD(-\xi)a^{\dag}(g) p_c(\xi),\end{split}
\end{equation}
and the photon-added state as
\begin{equation}
\begin{split}\label{eq:rhoExpandedAdd}
\rho^+ &= \frac{a^{\dag}(g)\rho_G a(g)}{\<\hat{n}(g)\>_G+1} \\ &= \frac{1}{\<\hat{n}(g)\>_G+1} \int {\rm d}^{2m}\xi \, a^{\dag}(g)D(\xi)\rho_sD(-\xi)a(g) p_c(\xi),\end{split}
\end{equation}
where we initially focus on the pure subtraction of a photon from mode $g \in {\cal N}(\mathbb{R}^{2m})$.
The evaluation of $a(g)D(\xi)\rho_sD(-\xi)a^{\dag}(g)$ is cumbersome and is therefore left for Appendix \ref{eq:SubAddDisp}, where we describe the general subtraction/addition of a photon from/to a displaced state. In general, we can use (\ref{eq:WignerDispApp}) to write the Wigner function of $\rho^{\pm}$ in (\ref{eq:rhoExpandedSub}) and (\ref{eq:rhoExpandedAdd}) as
\begin{equation}\begin{split}\label{eq:WignerExpandedSub}
W^{-}(\beta) = \int&{\rm d}^{2m}\xi \, W^{-}_{\xi}(\beta)\\&\times\frac{ \<\hat{n}(g)\>_s + \frac{1}{4}\big[ (\xi, g)^2 + (\xi, Jg)^2\big]}{\<\hat{n}(g)\>_G}p_c(\xi),
\end{split}
\end{equation}
and
\begin{equation}
\begin{split}\label{eq:WignerExpandedAdd}
W^{+}(\beta) = \int&{\rm d}^{2m}\xi \, W^{+}_{\xi}(\beta)\\&\times\frac{ \<\hat{n}(g)\>_s+1 + \frac{1}{4}\big[ (\xi, g)^2 + (\xi, Jg)^2\big]}{\<\hat{n}(g)\>_G+1}p_c(\xi),
\end{split}
\end{equation}
with
\begin{align}\label{eq:WignerDisp}
W^{\pm}_{\xi}(\beta)=  &\frac{W_{s}(\beta - \xi)}{\tr\big( (V_s + \norm{\xi}^2P_{\xi} \pm \mathbb{1})(P_{g}+P_{J\!g}) \big)} \\&\times\Bigg( \norm{(P_g+P_{J\!g})(\mathbb{1}\pm V_s^{-1})  (\beta-\xi)}^2\nonumber \\&\qquad+  2\big(\xi, (P_g+P_{J\!g})(\mathbb{1}\pm V_s^{-1}) (\beta-\xi) \big)\nonumber\\&\qquad+ \tr\big((P_g +P_{J\!g})(\norm{\xi}^2P_{\xi} - V_s^{-1} \mp \mathbb{1} )\big) \Bigg)\nonumber.
\end{align}
Note that $W_s$ and $V_s$ denote the Wigner function and covariance matrix, respectively, of $\rho_s$, as introduced in (\ref{eq:notoriousDecomp}, \ref{eq:GaussianMix}). The passive separability of the Wigner functions (\ref{eq:WignerExpandedSub}) and (\ref{eq:WignerExpandedAdd}) now depends on two aspects. Firstly, it hinges on the factorisability of the pure state Wigner function $W^{\pm}_{\xi}(\beta)$, as given by (\ref{eq:WignerDisp}). Secondly, we require that 
\begin{equation}
p^{-}_c(\xi) \equiv \frac{ \<\hat{n}(g)\>_s + \frac{1}{4}\big[ (\xi, g)^2 + (\xi, Jg)^2\big]}{\<\hat{n}(g)\>_G}p_c(\xi),
\end{equation}
and
\begin{equation}
p^{+}_c(\xi) \equiv \frac{ \<\hat{n}(g)\>_s+ 1 + \frac{1}{4}\big[ (\xi, g)^2 + (\xi, Jg)^2\big]}{\<\hat{n}(g)\>_G + 1}p_c(\xi),
\end{equation}
are well-defined probability distributions. \\

\paragraph*{Probability distributions $p^{\pm}_c(\xi)$---} It is straightforwardly verified that $p^{\pm}_c(\xi)$ are well-defined probability distributions. Because we know that $p^{-}_c$ and $p^{+}_c$ are positive functions, it suffices to validate their normalisation. To do so, we evaluate
\begin{equation}\begin{split}
\int&{\rm d}^{2m}\xi \, \big[ (\xi, g)^2 + (\xi, Jg)^2\big] p_c(\xi) \\
&= \int{\rm d}^{2m}\xi \, \big[ (\xi, g)^2 + (\xi, Jg)^2\big]  \frac{\exp\left\{-\frac{(\xi, V_c^{-1 }\xi)}{2}\right\}}{(2\pi)^m \sqrt{\det V_c}}\\
&= (g, V_c g) + (Jg, V_c Jg) = \tr \{(P_g+ P_{Jg}) V_c\} , 
\end{split}
\end{equation}
where we used that 
\begin{equation}
\int{\rm d}^{2m}\xi \, P_{\xi} \norm{\xi}^2 \frac{\exp\left\{-\frac{(\xi, V_c^{-1 }\xi)}{2}\right\}}{(2\pi)^m \sqrt{\det V_c}} = V_c.
\end{equation}
This implies that
\begin{equation}\begin{split}
\int{\rm d}^{2m}\xi \,p^{-}_c(\xi) &= \frac{ \<\hat{n}(g)\>_s + \frac{1}{4}\tr \{(P_g+ P_{Jg}) V_c\}}{\<\hat{n}(g)\>_G}\\
&= \frac{1}{4 \<\hat{n}(g)\>_G}\Big(\tr \{(P_g+ P_{Jg}) (V_s-\mathbb{1})\} 
\\&\qquad\qquad\qquad+\tr \{(P_g+ P_{Jg}) V_c\}\Big)\\
&= \frac{1}{4}\frac{  \tr \{(P_g+ P_{Jg}) (V-\mathbb{1})\}}{\<\hat{n}(g)\>_G} =1,
\end{split}
\end{equation}
where we use that, by construction, $V=V_s+V_c$. Analogously, we find that
\begin{equation}\begin{split}
\int{\rm d}^{2m}\xi \,p^{+}_c(\xi) & =1.
\end{split}
\end{equation}

\paragraph*{Factorisability of $W^{\pm}_{\xi}(\beta)$---} Because $p^{\pm}_c(\xi)$ are probability distributions, the states (\ref{eq:WignerExpandedSub}) and (\ref{eq:WignerExpandedAdd}) are passively separable whenever a mode basis exists in which $W^{\pm}_{\xi}(\beta)$ factorises for every $\xi$. The factor $W_{s}(\beta - \xi)$ in (\ref{eq:WignerDisp}) directly fixes a basis in which this problem must be considered because $W_{s}(\beta - \xi)$ only factorises in a product of single-mode Wigner functions in the symplectic basis of eigenvectors of $V_s$. %This basis can be obtained through the $O$ in the Bloch-Messiah decomposition (\ref{eq:splitV}). 
We denote this basis as ${\cal E}_s = \{e^{(1)}, \dots e^{(m)},J\!e^{(1)},\dots J\!e^{(m)}\}$ and decompose the vectors $\beta$ and $\xi$ as
\begin{align}\label{eq:BasisExpansion2} &\beta=\sum_{j=1}^m \beta^{(j)}_x e^{(j)}+ \beta^{(j)}_pJ\!e^{(j)}, \\ &\xi=\sum_{j=1}^m \xi^{(j)}_x e^{(j)}+ \xi^{(j)}_pJ\!e^{(j)},\end{align} 
from which it directly follows that, in this basis, the pure state Wigner function $W_s(\beta-\xi)$ takes the form
\begin{equation}
W_s(\beta-\xi) = \prod_{j=1}^{m}W^{(j)}_s\left(\beta^{(j)}_x-\xi^{(j)}_x, \beta^{(j)}_p-\xi^{(j)}_p\right).
\end{equation}
Next, we consider the behaviour of the polynomial
\begin{align}\label{eq:P2}
&P_2(\beta-\xi)\\&=\Bigg( \norm{(P_g+P_{J\!g})(\mathbb{1}\pm V^{-1})  (\beta-\xi)}^2 \nonumber \\&\qquad+  2\big(\xi, (P_g+P_{J\!g})(\mathbb{1}\pm V^{-1}) (\beta-\xi) \big)\nonumber\\&\qquad+ \tr\big((P_g +P_{J\!g})(\norm{\xi}^2P_{\xi} - V^{-1} \mp \mathbb{1} )\big) \Bigg),\nonumber
\end{align}
in this basis. The factorisability of $P_2(\beta-\xi)$ is completely governed by the vector $(P_g+P_{J\!g})(\mathbb{1}\pm V_s^{-1})(\beta-\xi)$. Because of the projector $(P_g+P_{J\!g})$, $P_2(\beta-\xi)$ is in essence a single-mode function determined by the mode $g$. 
It is straightforwardly verified that in case there exists a mode $i$ in the symplectic basis of eigenvectors of $V_s$ for which $g \in {\rm span}\{e^{(i)}, J\!e^{(i)}\}$, we find
\begin{equation}\begin{split}\label{eq:factorisationIsHere}
W^{\pm}_{\xi}(\beta)= &W^{(i)}_{\pm}\left(\beta^{(i)}_x-\xi^{(i)}_x, \beta^{(i)}_p-\xi^{(i)}_p\right)\\&\times \prod_{\substack{j=1\\ j \neq i}}^{m}W^{(j)}_s\left(\beta^{(j)}_x-\xi^{(j)}_x, \beta^{(j)}_p-\xi^{(j)}_p\right),
\end{split}
\end{equation}
such that $W^{\pm}_{\xi}$ factorises in the mode basis ${\cal E}_s$ for any displacement $\xi$.\\

We showed that the Wigner function of the single-photon added and subtracted states can always be represented as \begin{equation}
W^{\pm}(\beta) = \int{\rm d}^{2m}\xi \, W^{\pm}_{\xi}(\beta)  p^{\pm}_c(\xi).
\end{equation}
Whenever a photon is added to or subtracted from a mode which is part of a basis in which $\rho_s$ factorises there exists a basis in which $W^{\pm}_{\xi}$ factorises (\ref{eq:factorisationIsHere}) for any $\xi$. Thus, in this case, there exists a mode basis in which the Wigner function is of the form (\ref{eq:entGen}), because $p^{\pm}_c(\xi)$ is a well-defined probability distribution. In other words, subtracting (adding) a photon from (to) a mode which is part of a basis in which $\rho_s$ factorises leads to the intuitive result that the state remains passively separable.

It follows automatically from (\ref{eq:photonSubRho}) and (\ref{eq:photonAddRho}) that we can generalise this approach to the scenario where the subtraction or addition process is not pure. In this case we define \begin{equation}\lambda_k = \frac{\gamma_k \tr\{(V \pm \mathbb{1})(P_{g_k} + P_{Jg_k})\}}{\tr\{ (V \pm \mathbb{1})\sum_k \gamma_k (P_{g_k} + P_{Jg_k})\}},\end{equation} and find that
\begin{align}
W^{\pm}_{\rm mix} (\beta) &= \sum_{k} \lambda_k  W^{\pm}_{g_k}(\beta) \\
&=  \int{\rm d}^{2m}\xi \, p^{\pm}_c(\xi) \sum_{k} \lambda_k W^{\pm}_{g_k,\xi}(\beta) \label{eq:mixSeparable} ,
\end{align}
where $W^{\pm}_{g_k}(\beta)$ is the Wigner function of (\ref{eq:WignerExpandedSub}) or (\ref{eq:WignerExpandedAdd}) for a specific subtraction mode $g_k$, and analogously $W^{\pm}_{g_k,\xi}(\beta)$ is given by (\ref{eq:WignerDisp}) for a specific subtraction (addition) mode $g_k$. Thus, if one can find a set of subtraction (addition) modes $\{g_k\} $, such that for any of these modes $g_k \in {\rm span}\{e^{(k)}, J\!e^{(k)}\}$, with $e^{(k)}, J\!e^{(k)} \in {\cal E}_s$, the state is passively separable.

Every $V_s \leqslant V$ gives rise to a possible decomposition (\ref{eq:thisdecomphere}) of the Gaussian state's covariance matrix $V$. Hence, each of these possible $V_s$ leads to a different basis ${\cal E}_s$ of symplectic eigenvectors with an associated set of eigenmodes. Subtracting or adding the photon in any mode in any such basis will leave the final state passively separable. In other words, {\em the state is passively separable whenever the photon is subtracted or added in a mode which is part of any mode basis for which the initial Gaussian state is separable}. It is, on the other hand, unclear that subtracting or adding a photon in any other mode automatically induces inherent entanglement. The reason is that we must consider all possible decompositions of the state $\rho^{\pm}$ in convex combinations of pure states. A priori, it is possible that convex combinations exist, which are not of the form (\ref{eq:WignerExpandedSub}, \ref{eq:WignerExpandedAdd}). Also for such decompositions linear separability must be excluded to prove inherent entanglement. This issue falls outside of the scope of our present work and is left as an open problem.\\

%In the special case of pure states, the discussion is significantly simplified. In this case the linear separability of the state $\rho^{\pm}$ depends only on the existence of a symplectic basis for which $W^{\pm}_{\xi}(\beta)$ in (\ref{eq:WignerDisp}) factorises. The latter describes the general Wigner function which is obtained when we subtract (or add) a photon from (or to) a displaced Gaussian state, and for pure state this Wigner function cannot be written as a convex combination of other Wigner functions. 

In the special case of a pure state $V=V_s$ is a symplectic matrix, such that we can directly find the mode basis ${\cal E}_s$ where $W_{s}(\beta - \xi)$ in (\ref{eq:WignerDisp}) factorises. If there is no mode $i$ for which $g \in {\rm span}\{e^{(i)}, J\!e^{(i)}\}$, we find that $P_2(\beta-\xi)$ in (\ref{eq:P2}) is a sum of terms associated with different modes of the basis ${\cal E}_s$. Because $W^{\pm}_{\xi}(\beta)$ in (\ref{eq:WignerDisp}) is the Wigner function of a pure state, it is impossible to write it as statistical mixture of Wigner functions. Moreover, we cannot factorise $W^{\pm}_{\xi}(\beta)$ in the basis where $W_{s}(\beta - \xi)$ is factorised. However, in any other basis there are cross-terms in $W_{s}(\beta - \xi)$ which prevent its factorisations and are associated with off-diagonal terms in $V_s$ which correlate different modes. These multimode factors in $W_{s}(\beta - \xi)$ can never be compensated by terms in $P_2(\beta-\xi)$. Therefore the state can never be separable in a mode basis where $W_{s}(\beta - \xi)$ does not factorise. This implies that for photon-added and -subtracted pure states the state is passively separable {\em if and only if} the subtraction or addition takes place in a mode from the mode basis for which the initial Gaussian state is separable. This mode basis coincides with the modes obtained from the Bloch-Messiah decomposition, commonly referred to as {\em supermodes}.

We stress that this implies that subtracting a photon from (or adding it to) a pure Gaussian state in a superposition of supermodes will always induce entanglement. Moreover, this entanglement is robust against linear operations in the sense that it cannot be undone by passive linear optics. Therefore, this type of inherent entanglement is clearly different from the Gaussian entanglement discussed in Section \ref{sec:Entanglement}.

%Because this result holds for a wide range of pure states, it fuels our conjecture that an extension to mixed states must be possible.

%The decomposition (\ref{eq:notoriousDecomp}) leads us to a unique (assuming no degeneracies in the squeezing) set of locally squeezed supermodes on which additional classical noise is added. Through this decomposition, we have shown that one can write any single photon added or subtracted state as a mixtures of single photon added or subtracted pure states. Natural is these steps may seem, it is far from obvious that no other decomposition of $\rho^{\pm}$ as a mixture of pure states fulfils (\ref{eq:entGen}).

%This point can be tackled through several results related to extremality of entanglement of Gaussian states. A covariance matrix, built up by the two-point correlations, can be associated to any possible state. A priori, many states may give rise to the same covariance matrix. However, it can be shown that for all these states associated to a specific covariance matrix, the Gaussian state which is fully characterised by it will have the lowest entanglement. In other words, if a Gaussian state with a particular covariance matrix is entangled, all other states with the same covariance matrix will also be entangled.

%For photon-added or -subtracted states, the complete covariance matrix is given by (\ref{eq:BoomAmix}) $V + A^{\pm}_{\rm mix}$. The Gaussian state characterised by $V + A^{\pm}_{\rm mix}$ is simply the initial Gaussian state (characterised by $V$) to which classical noise (characterised by $A^{\pm}_{\rm mix}$) was added. 

\subsection{Algebraic interpretation}

The creation of inherent entanglement due to single-photon addition and subtraction in the pure state case can also be understood in an algebraic way. To do so, we define the operator ${\cal O}$ on the Hilbert space that describes the system's states, which implements a change in mode basis. On the mode space, this basis change can be implemented by the orthogonal symplectic matrix $O$. In other words, ${\cal O}$ describes a linear optics circuit. The action of ${\cal O}$ on the quadrature operators is given by
\begin{align}
{\cal O}^{\dag} Q(f) {\cal O} = Q(Of),
\end{align}
such that the structure of the canonical commutation relations (\ref{eq:ccr}) remains conserved. Because we demand $O$ to be a symplectic matrix, it follows that $OJ = JO$. The definitions (\ref{eq:creationAnni}) of the creation and annihilation operators then imply that
\begin{equation}
{\cal O}^{\dag} a(g) {\cal O} = a(Og), \quad \text{and} \quad {\cal O}^{\dag} a^{\dag}(f) {\cal O} = a^{\dag}(Og).
\end{equation}
This implies also that we can write any photon-subtraction given by $a(g)$ as a photon subtraction in a different mode with additional linear optic operations, since $a(g) = {\cal O}^{\dag} a(O^t g) {\cal O}$. The operation ${\cal O}$ acts in a very natural way on a Gaussian state $\rho_G$, with covariance matrix $V$. Indeed, we find that ${\cal O}\rho_G {\cal O}^{\dag} = \rho'_G$, which is a Gaussian state with covariance matrix $O V O^t$. 
%This directly highlights the potential of ${\cal O}$ to induce Gaussian entanglement between modes.

\begin{figure}
\centering
\includegraphics[width=0.5\textwidth]{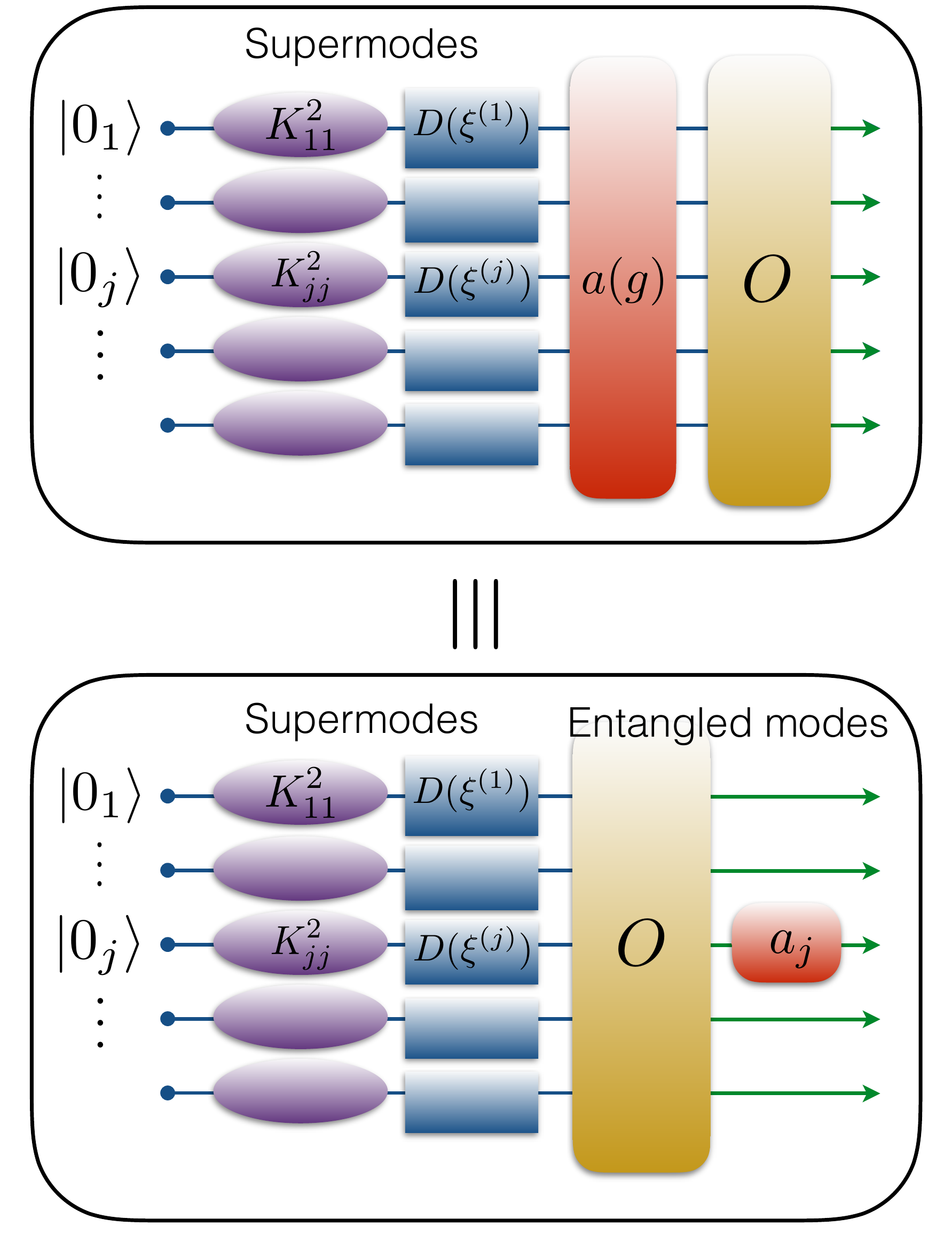}
\caption{Circuit representation of eqs.~(\ref{eq:bleeb}-\ref{eq:bloob}) for a pure initial state $\rho_G$, represented in the mode basis obtained by the Bloch-Messiah decomposition (blue lines) as a vacuum ($\ket{0_1} \otimes \dots \otimes \ket{0_m}$) which is locally squeezed ($K^2_{jj}$) and displaced $D(\xi^{(j)})$. The top panel shows general photon-subtraction is implemented by the action of annihilation $a(g)$ in mode $g \in {\cal N}(\mathbb{R}^{2m})$, which is nonlocal (see main text) in the supermode basis. The mode basis is changed by the action of a linear optics operation (\ref{eq:bleeb}), characterised by orthogonal symplectic matrix $O$. The bottom panel shows the equivalent procedure (\ref{eq:bloob}), where the mode basis is changed (green lines) before subtracting the photon. Here we assume that $O^tg = e'^{(j)}$, which is on of the modes in the new mode basis (green lines), such that the subtraction is local. We denote $a_j = a(e'^{(j)})$.  \label{fig:Circuits}}
\end{figure}

When we now consider the action of a linear optics operations ${\cal O}$ on the state obtained through subtracting a photon, we find
\begin{align}\label{eq:bleeb}
{\cal O}\rho^{-}{\cal O}^{\dag} &=\frac{1}{\<\hat{n}(g)\>_G} {\cal O}a(g)\rho_Ga^{\dag}(g){\cal O}^{\dag}\\
& = \frac{1}{\<\hat{n}(g)\>_G}  a(O^tg){\cal O}\rho_G{\cal O}^{\dag}a^{\dag}(O^tg)\\
&= \frac{1}{\<\hat{n}(g)\>_G}  a(O^tg)\rho'_Ga^{\dag}(O^tg).\label{eq:bloob}
\end{align}
In practice, these equalities describe very different way of preparing an identical state as seen on Fig.~\ref{fig:Circuits}. The discussion for photon-addition is identical. 

%If we now assume that we treat the problem initially in the mode basis ${\cal E}_s = \{e^{(1)}, \dots e^{(m)},J\!e^{(1)},\dots J\!e^{(m)}\}$ where a pure state $\rho_G$ is separable, and thus its covariance matrix $V$ is diagonal, we can use this approach to understand entanglement properties. When a photon is added or subtracted in a mode $g \in {\cal N}(\mathbb{R}^{2m})$ which is not part of ${\cal E}_s$, we can use a linear optics transformation to localise the photon. This operation effectively conducts a basis change to a new mode basis ${\cal E}'_s = \{e'^{(1)}, \dots e'^{(m)},J\!e'^{(1)},\dots J\!e'^{(m)}\}$, such that any ${\cal O}$ which generates a mode transformation $O$ for which $O^t g \in {\rm span}\{e'^{(i)}, J\!e'^{{i}}\}$ localises the subtraction process to a mode $i$ in the new basis. However, it also transforms the covariance matrix of $\rho_G$ as $V \mapsto OVO^t$. Unless there are degeneracies in $V$, this operation will always induce Gaussian entanglement between the modes in $\rho'_G$ (\ref{eq:gaussTransLinOpt}). In other words, initially, all entanglement in the photon-subtracted state was strictly non-Gaussian in nature, induced by non-local and non-Gaussian operation on  a separable Gaussian state. Using linear optics, we can localise the non-Gaussian operation while also performing an entangling operation on the initial Gaussian state. Hence, linear optics allows us to ``exchange'' Gaussian and purely non-Gaussian entanglement, but entanglement as such will always remain present. 

The action of linear optics (\ref{eq:bleeb} - \ref{eq:bloob}) in the case where $O^t g = e'^{(j)}$ --such that the transformation localises the photon subtraction to an entangled mode-- can be represented in a more graphic way, using a type of circuit representation. Specifically, in Fig.~\ref{fig:Circuits} we show how all the different operations act with respect to the modes where the squeezing is local (as obtained by the Bloch-Messiah decomposition). It highlights clearly that we can always find a mode basis where squeezing, displacements and the photon-subtraction (or addition) act {\em locally}. Even though experimentally one can consider co-propagating modes \cite{ra_tomography_2017}, it is always possible to spatially separate these different degrees of freedom. Hence the term ``local'' can be physically understood in this sense.

%All entanglement is induced by a linear optics transformation between the preparation of the Gaussian state and the photon-subtraction in one of these modes.

\subsection{Reduced states}\label{sec:Red}

An important tool to measure entanglement in quantum systems is the reduced quantum state. When we study entanglement, we typically fix a partition of the system to entangle (in the case of multimode quantum optics, this partition is comprised of different modes). The reduced state is obtained by integrating (or ``tracing'') out several of these degrees of freedom. These reduced states are important in the study of entanglement properties, specifically when the full state is pure.

The methods provided in Section \ref{sec:phaseSpace} are ideally suited to derive the Wigner functions for the reduced states of a multimode photon-added or -subtracted state. In particular, we stress that the characteristic function (\ref{eq:Chi}) is in principle obtained in a single mode fashion. Therefore, we can obtain the characteristic function of the reduced state, associated with a mode space ${\cal M} \subset {\cal N}(\mathbb{R}^{2m})$, by simply restricting $\alpha$ in (\ref{eq:Chi}) to $\alpha = \lambda f$ with $f \in {\cal M}$ and $\lambda \in \mathbb{R}$. 

We can now define a symplectic basis ${\cal E}_{\cal M} = \{\nu^{(1)}, \dots, \nu^{(m')}, J\!\nu^{(1)}, \dots, J\!\nu^{(m')} \}$ of ${\cal M}$, where $\dim {\cal M} = m' < m$. The restrictions of the matrices $V$ and $A^{\pm}_{{\rm mix}}$ to ${\cal M}$ are denoted by $V^{\cal M}$ and $A_{{\rm mix}}^{\cal M}$, respectively. These restricted matrices are the ones which are obtained by only measuring the correlations among the modes in ${\cal M}$, which directly follows from (\ref{eq:BoomAmix}) and (\ref{eq:TruncFinalMix}). This straightforwardly implies that the Wigner function of the state on the reduced mode set ${\cal M}$ has the form
\begin{align}
&W^{\pm}_{\cal M}(\beta') \label{eq:redWigner}\\
&\quad= \frac{1}{ 2} \Big((\beta', {V^{\cal M}}^{-1} A^{{\cal M}}_{\rm mix} {V^{\cal M}}^{-1} \beta') -  \tr({V^{\cal M}}^{-1}A^{\cal M}_{\rm mix}) + 2 \Big)\nonumber\\
&\qquad\qquad\times  \frac{1}{(2\pi)^m \sqrt{\det V^{\cal M}}} e^{-\frac{1}{2}(\beta', {V^{\cal M}}^{-1} \beta')},\nonumber
\end{align}
where $\beta'$ is a vector in the optical phase space $\beta' \in \mathbb{R}^{2m'}$ associated with ${\cal M}$.

The Wigner function of the reduced state naturally provides us with a measure for the entanglement of the state if the initial $m$-mode state is pure. Indeed, selecting a set of modes determines a bipartition which contains the modes ${\cal M}$ on the one hand, and the modes which were integrated out, ${\cal M}^{\perp}$, on the other hand. The purity of the reduced state on the modes ${\cal M}$ is directly related to the entanglement between the modes in ${\cal M}$ and in ${\cal M}^{\perp}$. The less pure the reduced state, the higher the entanglement between the two parts of the bipartition. 

Once the Wigner function of the reduced state is obtained, its purity can directly be evaluated. It is given by
\begin{equation}\label{eq:absPurity}
\mu_{\cal M} = (4\pi)^{m'}\int_{\mathbb{R}^{2m'}} {\rm d}^{2m'}\beta' \big(W^{\pm}_{\cal M}(\beta')\big)^2.
\end{equation}
In the following section, we will use (\ref{eq:absPurity}) to evaluate the pure state results obtained in Section \ref{sec:EntanglementAddSub}.

\section{Examples}\label{sec:examples}

In the above sections, we developed a framework for the analysis of single-photon added and subtracted states. The goal of this section is to provide two examples to highlight the usefulness of the above results. In the first example, we treat the well-known case of a two-mode squeezed vacuum, where both modes are equally squeezed. The second example extends the study in \cite{walschaers_entanglement_2017}, and uses an initial Gaussian state which was experimentally obtained \cite{cai_reconfigurable_2016}.

\subsection{Two-mode symmetrically squeezed vacuum}\label{sec:ex1}

The two-mode symmetrically squeezed vacuum state is in a certain sense the simplest nonclassical multimode Gaussian state. We characterise it through its covariance matrix $V_s$, which takes the form
\begin{equation}\label{eq:VtwoMode}
V_s =\begin{pmatrix}10^{-s/10} & 0&0 &0 \\ 0 &10^{-s/10} & 0&0 \\ 0  & 0&10^{s/10}&0 \\ 0& 0&0&10^{s/10},  \end{pmatrix}
\end{equation}
in its basis of eigenmodes. The notation is chosen such that $s$ denotes the amount of squeezing in dB.\\

At first, we use the results presented in Sections \ref{sec:phaseSpace} and \ref{sec:Red} to investigate the entanglement properties which are induced by adding or subtracting a photon. With the covariance matrix (\ref{eq:VtwoMode}) we have all the necessary information to construct the Wigner function (\ref{eq:WignerFinalElegant}) that describes a state with a single photon added or subtracted in mode $g \in {\cal N}(\mathbb{R}^4)$. To induce entanglement, we must subtract a photon in a superposition of two supermodes. Due to the symmetry properties of the state it is also important to include a phase in this superposition. Hence, the most interesting choice is $g = (x_1,0,0,x_2)^t$ with $x_ 1^2+x_2^2 = 1$.

\begin{figure*}
\centering
\includegraphics[width=0.99\textwidth]{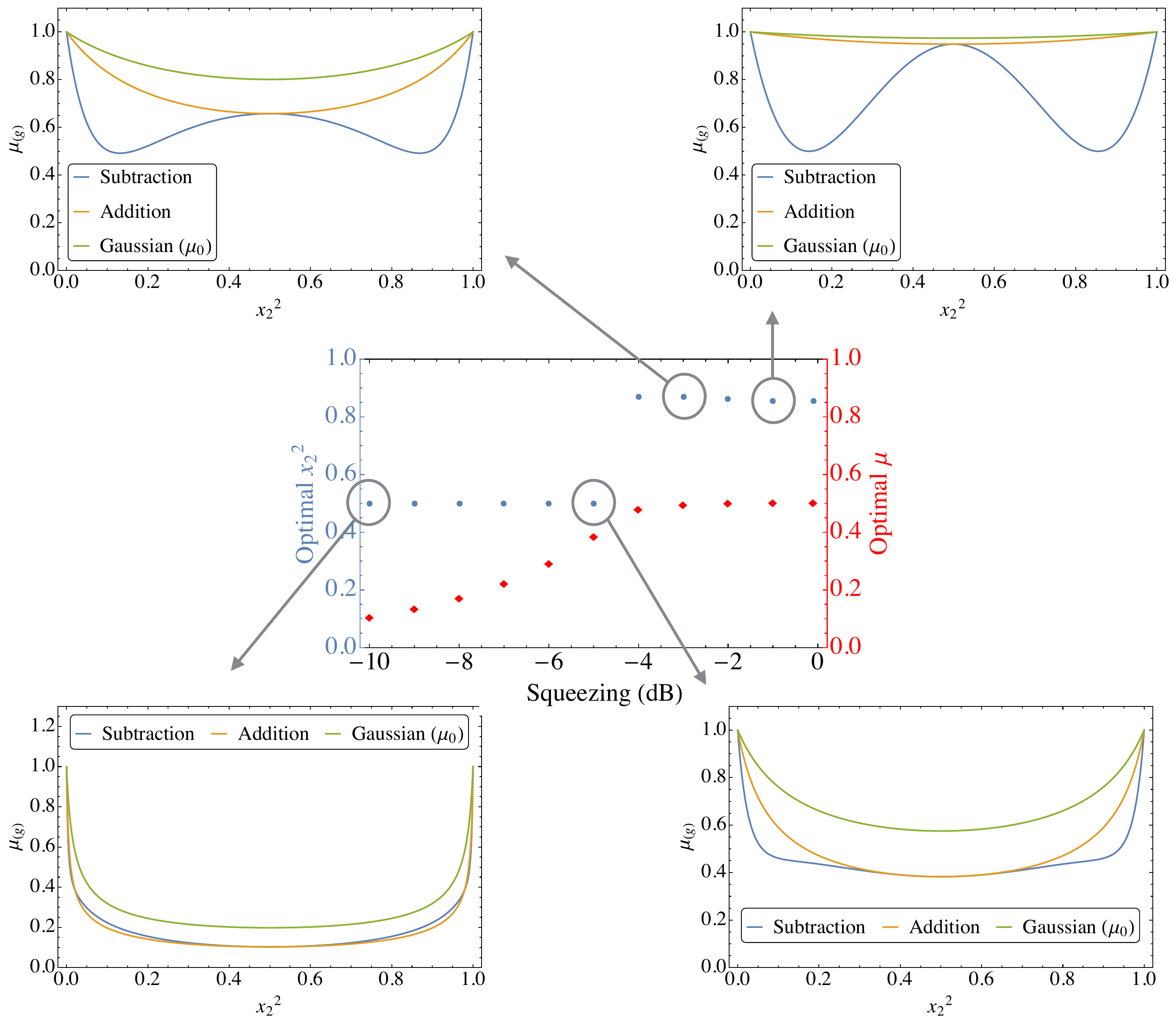}
\caption{{\bf Exterior panels:} Purity (\ref{eq:absPurity}) of the state reduced to the mode $g = (x_1,0,0,x_2)^t$, in which the photon was added (orange) or subtracted (blue), as a function of the weight $x_2^2$ in the superposition of supermodes. The degree of squeezing, i.e.~$s$ in (\ref{eq:VtwoMode}) is set to $s=1{\rm dB}$ (top right), $s=3{\rm dB}$ (top left), $s=5{\rm dB}$ (bottom right) and $s=10{\rm dB}$ (bottom left). Purity of the reduced Gaussian state, in the same mode, is shown as a reference (green). Lower purity is directly related to higher entanglement between $g$ and the complementary orthogonal mode.\\
{\bf Centre panel:} Values of the weight $x_2^2$ (blue), with $g = (x_1,0,0,x_2)^t$, for which the lowest purity (red) --and thus the highest entanglement-- is achieved for different values of squeezing in the case of photon subtraction. Points associated with exterior panels are highlighted.\label{fig:TwoModeEnt}}
\end{figure*}

In Fig.~\ref{fig:TwoModeEnt}, we investigate the purity, $\mu$, (\ref{eq:absPurity}) for the reduce density matrix associated with the mode where the photon is subtracted or added. This boils down to integrating out the mode orthogonal to $g$, or setting ${\cal M} = {\rm span}\{g, J\! g\}$ in (\ref{eq:absPurity}). In practice, we analytically construct the reduced state's Wigner function (\ref{eq:redWigner}) and perform a numerical integration to obtain the purity, denoted $\mu_{(g)}$. Because the purity of the reduce state is directly related to an entanglement measure if the initial state is pure, we can directly associate $\mu_{(g)}$ to the entanglement between $g$ and the complementary mode. Fig.~\ref{fig:TwoModeEnt} investigates what happens to the entanglement as we vary the only parameter left in the system: $x_2/x_1$, which governs $g$, and the squeezing $s$ in (\ref{eq:VtwoMode}). 

As a reference, let us first consider entanglement in the Gaussian state before the subtraction or addition has taken place. We observe entanglement between the mode ${\cal M} = {\rm span}\{g, J\! g\}$ and the complementary mode ${\cal M}^{\perp}$ in the initial squeezed vacuum state, characterised by (\ref{eq:VtwoMode}). It is clearly seen in Fig.~\ref{fig:TwoModeEnt} (green curves) that in the Gaussian state entanglement is maximal in the balanced case, i.e.~for $x_2^2 = 1/2$. Moreover, we observe that this Gaussian entanglement increases with increased squeezing. The results for photon addition (orange curves) are qualitatively the same as the Gaussian case. However, quantitatively the obtained purities are lower, implying that the addition of a photon increases entanglement.  

The results for photon subtraction in Fig.~\ref{fig:TwoModeEnt} (blue curves and central panel) are more surprising. For high squeezing, $s > 5 {\rm dB}$, we observe a similarity with the results for photon addition. However, for low squeezing, $s < 5 {\rm dB}$, the lowest purity is no longer obtained for the balanced superposition of supermodes $x_2^2 = 1/2$, but rather for the highly imbalanced superposition with $x_2^2 \approx 0.85$ \footnote{The exact value of $x_2^2$ for which the minimal purity is achieved weakly varies with the squeezing.}. Due to the symmetry in the squeezing, the system is completely unchanged when $x_1$ and $x_2$ are interchanged. This implies that exactly the same value for the purity is obtained for $x_2^2 \approx 0.15$. It is particularly surprising that, even for $s = 1$ in (\ref{eq:VtwoMode}), we still observe that $\mu_{(g)} = 0.5$ when $x_2^2 \approx 0.85$. Hence, we highlight a profound difference in the induced entanglement properties for the subtraction as compared to the addition of a photon. 

In the limit of low squeezing (i.e.~$s\rightarrow 0$) it is useful to analyse the state which is obtained by photon-subtraction in the photon representation. We straightforwardly obtain that, in the low squeezing limit, the state, denoted $\ket{\psi^-}$ is given by
\begin{equation}
\ket{\psi^-} = \frac{2(x_1^2 - x_ 2^2)\ket{1,0} + 4 x_1x_2 \ket{0,1}}{\sqrt{4(x_1^2 - x_ 2^2)^2 + 16 x_1^2x_2^2}},
\end{equation}
where $\ket{n_1,n_2}$ is the state with $n_1$ photon in mode one and $n_2$ photons in mode two. It is directly verified that for any solution with $x_1^2 = 1 - x^2_2$ and $x_2^2= (2 \pm \sqrt{2})/4$, the state $\ket{\psi^-}$ is a single-photon Bell state. This immediately gives the reason why we observe a reduced state purity of $1/2$ for these modes. The most remarkable aspect of our results, is that we show how this observed entanglement between the modes survives under reasonably high amounts of squeezing, where both modes are populated with many photons. Experimentally, this should make this phenomenon easier to observe.\\

\begin{figure}
\centering
\includegraphics[width=0.5\textwidth]{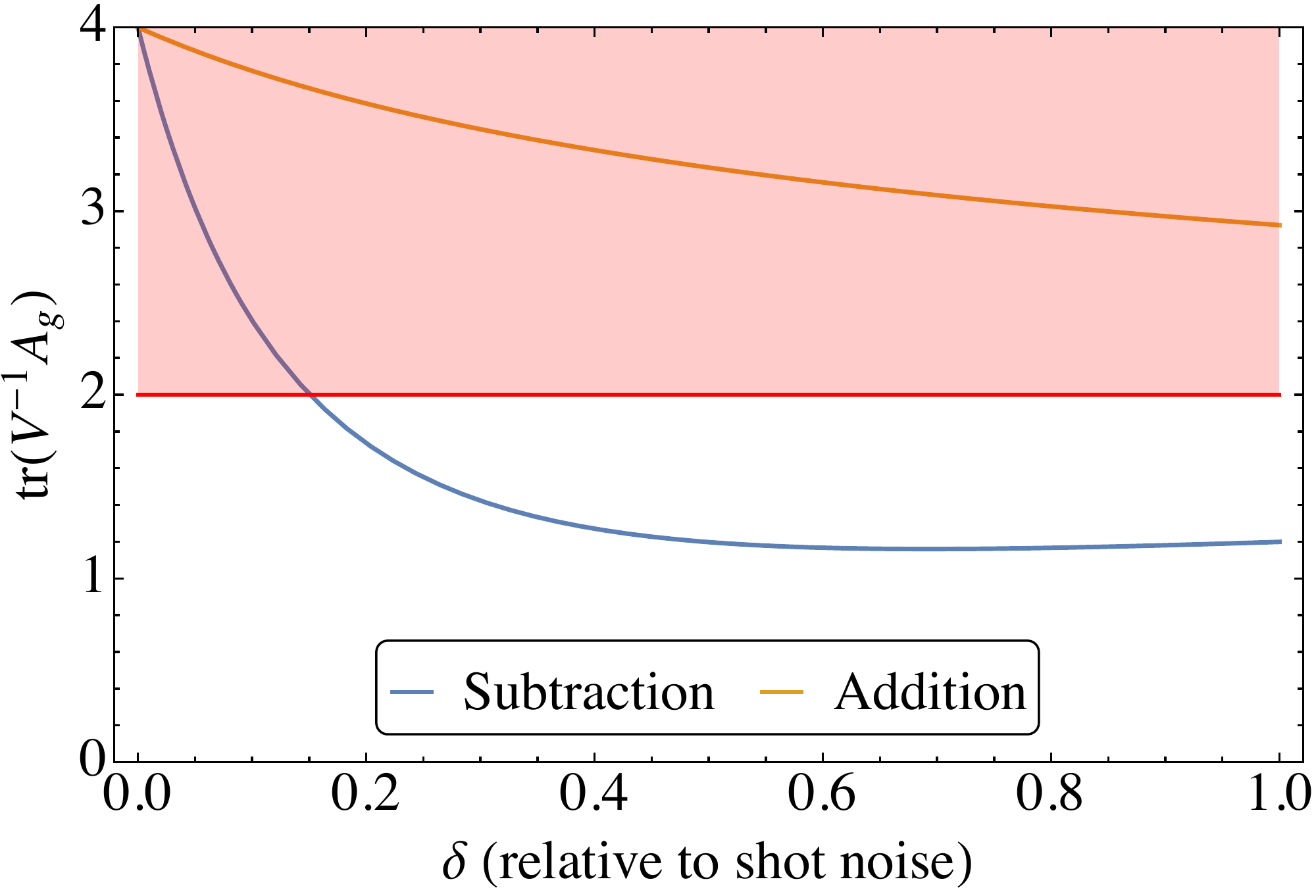}
\includegraphics[width=0.5\textwidth]{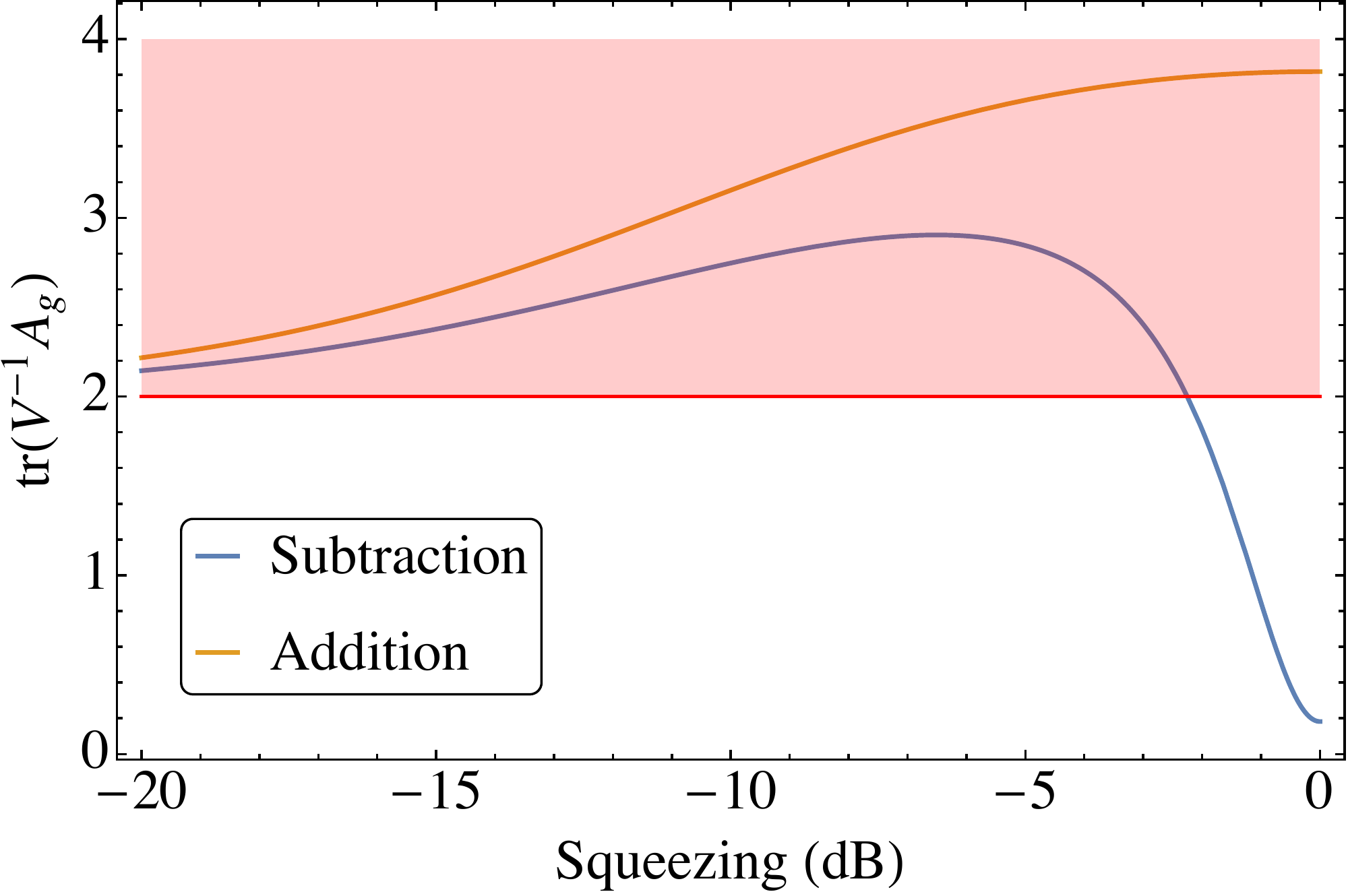}
\caption{The negativity condition (\ref{eq:conditionNegat}, red region) is shown as a function of the classical noise $\delta$ (\ref{eq:delta}) for a fixed squeezing of $3 {\rm dB}$ (top), and as a function of squeezing for a fixed noised of $\delta = 0.1$ as a fraction of the shot noise (bottom). The mode in which addition and subtraction take place is set to $g = (1/\sqrt{2},0,0,1/\sqrt{2})^t$.  \label{fig:TwoModeNeg}}
\end{figure}

Not only the entanglement properties, but also the negativity of the Wigner function behaves very differently for photon-added and -subtracted states. We probe the state's negativity through the witness (\ref{eq:conditionNegat}) and immediately observe that, both for addition and subtraction of a photon, $\tr(V^{-1}A^{\pm}_{g}) = 4$ when the state is pure. Hence, the condition (\ref{eq:conditionNegat}) for negativity is satisfied such that the Wigner function is negative. More interesting is the more general case of addition or subtraction from a mixed state. We approach this scenario through a simple noise model by considering an initial Gaussian state with a covariance matrix
\begin{equation}\label{eq:delta}
V = V_s + \delta \mathbb{1},
\end{equation}
where $\delta$ indicates the amount of added classical noise relative to the shot noise. In Fig.~\ref{fig:TwoModeNeg} we show how the negativity witness $\tr(V^{-1}A^{\pm}_{g})$ is influenced by the variation of the noise $\delta$ and of the squeezing $s$ in (\ref{eq:VtwoMode}). As derived in (\ref{eq:thiscoolone}), we observe that (\ref{eq:conditionNegat}) is always fulfilled for photon addition. In contrast, the negativity of the Wigner function for photon subtraction is very sensitive to the added noise $\delta$. It is not surprising that states which are more strongly squeezed are more robust to noise. Finally, we note that, due to the symmetry of the squeezing in both supermodes, $\tr(V^{-1}A^{\pm}_{g})$ is fully independent of the mode (or mixture of modes) in which the photon is added or subtracted.\\

In summary, we highlighted the potential of the methods of Section \ref{sec:phaseSpace} to study the photon addition and subtraction from a symmetric two mode squeezed vacuum. We showed that pure photon-subtracted states have highly interesting entanglement properties in the regime of low squeezing. Nevertheless, the negativity of the Wigner function --which is a crucial property to reach a quantum advantage in computation-- is much more sensitive to noise for photon subtraction than for photon addition. A detailed understanding of the interplay of these negativities and the entanglement properties of the states lies beyond the scope of this work.

\subsection{Experimentally generated Gaussian state}

In \cite{walschaers_entanglement_2017} we already presented results for an initial Gaussian state which was experimentally obtained \cite{cai_reconfigurable_2016}. Here, we complement these results with, on the one hand, additional findings for the entanglement between modes in the pure state part. On the other hand, we provide a study of the effect of impure addition and subtraction (in the sense of (\ref{eq:photonSubRhoMix}) and (\ref{eq:photonAddRhoMix})) on the negativity of the resulting Wigner function.\\

\paragraph*{Inherent entanglement---}We restrict our study of inherent entanglement is restricted to pure states, in accordance with Section \ref{sec:EntanglementAddSub}. However, the covariance matrix $V$ of  \cite{cai_reconfigurable_2016} is not symplectic, i.e.~we cannot find a symplectic basis of eigenvectors, which directly implies that the state cannot be pure. This can explicitly be seen in the Williamson decomposition $V = S^t \Delta S$, where $S$ is symplectic and $\Delta \geqslant \mathbb{1}$ a diagonal matrix. For a pure state, one must find that $\Delta = \mathbb{1}$, which is not the case for $V$. Hence, we will investigate entanglement properties for photon addition and subtraction from a pure squeezed vacuum which is consistent with $V$. 
%A priori, as stressed in Section \ref{sec:Entanglement}, any symplectic covariance matrix $V_s \leqslant V$ can be considered a valid choice in this context. Nevertheless, 
To do so, we resort to the Bloch-Messiah decomposition $S = O' K O$, where $O$ and $O'$ are orthogonal symplectic matrices, and $K$ is a positive diagonal symplectic matrix. We then obtain that $V = O^t K O'^t  \Delta O' K O$, which we use to decompose $V= V_s + V_c$ as in Section \ref{sec:Entanglement}. Through this method, we obtained a squeezed vacuum which is characterised by $V_s = K^2$ in the basis of eigenmodes.

\begin{figure}
\centering
\includegraphics[width=0.49\textwidth]{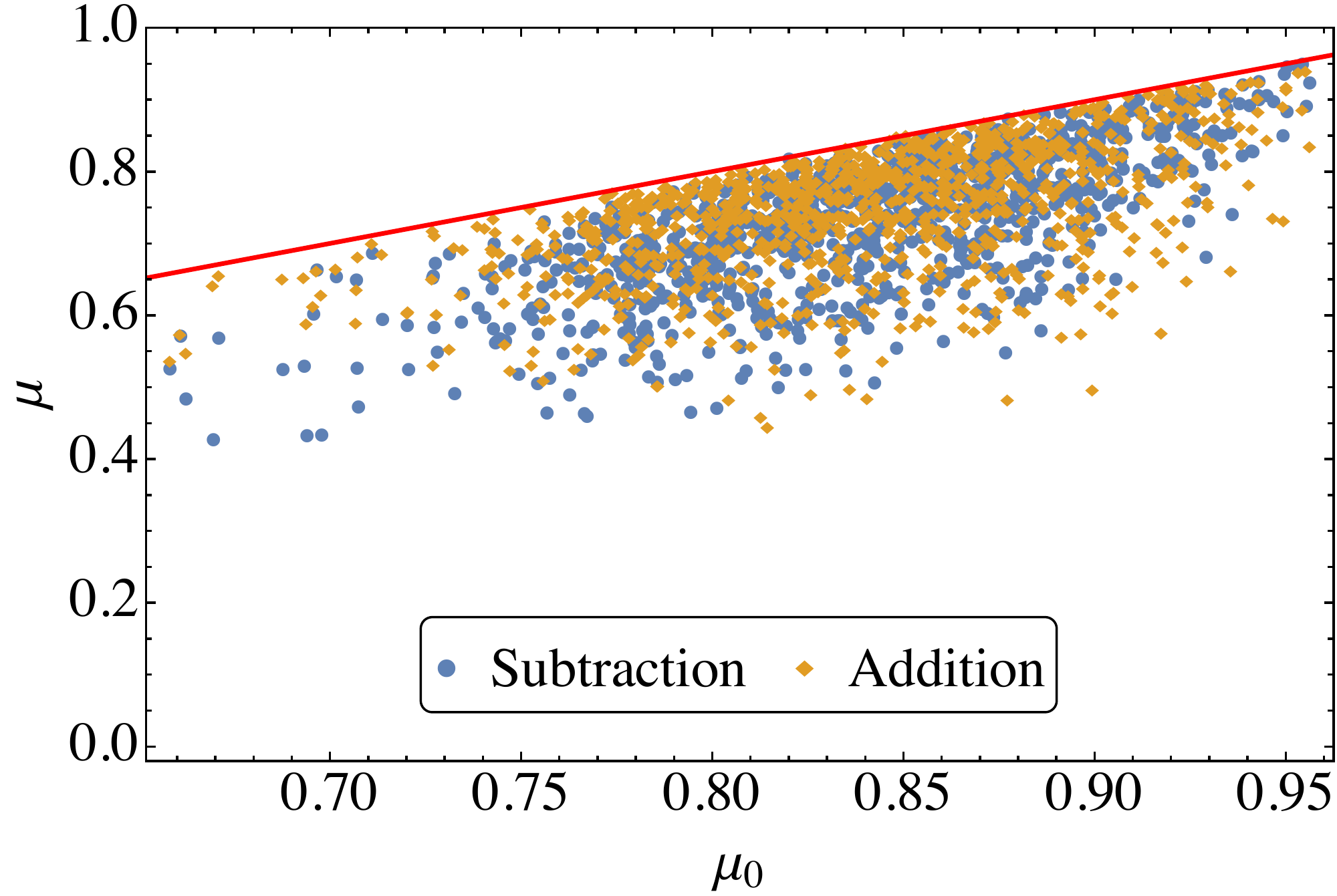}
\caption{Purity $\mu$ (\ref{eq:absPurity}) of the single-mode reduced state after the addition (orange diamonds) or subtraction (blue dots) of a photon from a pure state deduced (see main text) from the experimental state \cite{cai_reconfigurable_2016}. Purities $\mu$ are compared to the purity $\mu_0$ of the same mode's reduced state before the addition/subtraction, i.e.~the initial Gaussian state. For all realisations, the photon is added or subtracted in the same randomly chosen mode $g \in {\cal N}(\mathbb{R}^{32})$. The mode $f \in {\cal N}(\mathbb{R}^{32})$ to which the state is reduced is chosen randomly for each realisation. Points where $\mu = \mu_0$ are indicated by the red curve.\label{fig:ExpEnt1}}
\end{figure}

\begin{figure}
\centering
\includegraphics[width=0.49\textwidth]{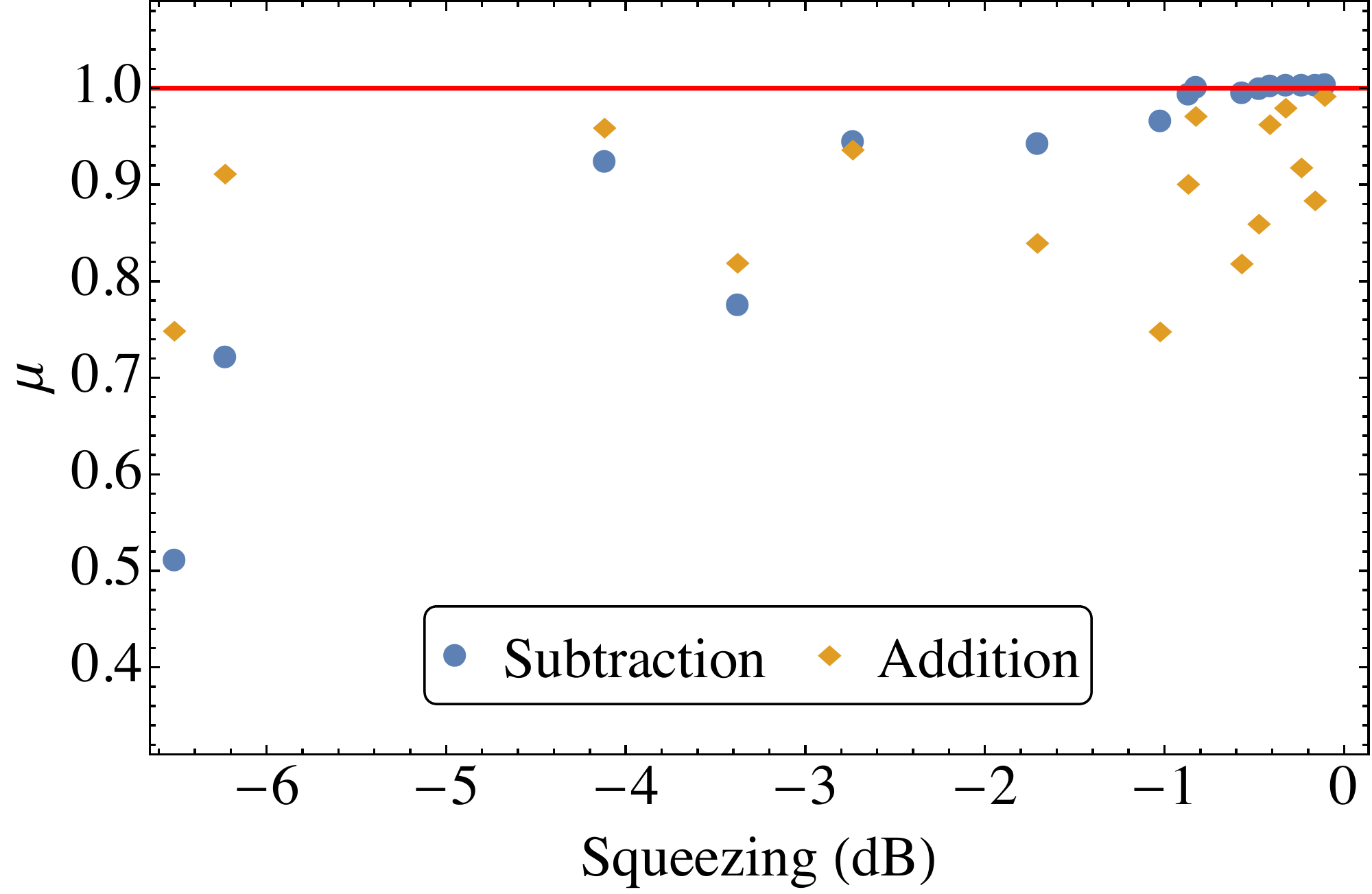}
\caption{Purity $\mu$ (\ref{eq:absPurity}) of the single-mode reduced state after the addition (orange diamonds) or subtraction (blue dots) of a photon in a pure state deduced (see main text) from the experimental state \cite{cai_reconfigurable_2016}. Each datapoint represents the reduction of the state to a different eigenmode of the initial Gaussian state's covariance matrix. The horizontal axis indicated the squeezing (in dB) of each of these eigenmodes. For all datapoints, the photon is added or subtracted in the same randomly chosen mode $g \in {\cal N}(\mathbb{R}^{32})$. The initial Gaussian state is fully separable the considered mode basis, hence the red curve indicates $\mu = 1$, in analogy with the red curve in Fig.~\ref{fig:ExpEnt1}.\label{fig:ExpEnt2}}
\end{figure}

Our previous work \cite{walschaers_entanglement_2017} showed that photon addition or subtraction can increase entanglement between the mode in which the photon is added or subtracted and the additional modes. The results in Section \ref{sec:EntanglementAddSub} show, moreover, that the subtraction or addition of a photon in a superposition of eigenmodes of $V_s$ induces entanglement in every possible mode basis. Here, in Figs.~\ref{fig:ExpEnt1} and \ref{fig:ExpEnt2}, we illustrate this point via the reduced state purity as a probe of the entanglement in specific bi-partitions. 

In both Figs.~\ref{fig:ExpEnt1} and \ref{fig:ExpEnt2}, a photon is subtracted from or added to the squeezed vacuum a random mode $g\in{\cal N}(\mathbb{R}^{2m})$. First, we subtract or add a photon to the squeezed vacuum, and reduce the obtained state to a randomly chosen mode $f$, as in Section \ref{sec:Red}. The purity $\mu$ of these reduced photon-added and -subtracted states is compared to the purity $\mu_0$ of the reduced squeezed vacuum prior to the addition or subtraction of a photon for the same mode $f$. By probing $1,000$ different random choices for $f$, Fig.~\ref{fig:ExpEnt1} shows entanglement for every bi-partition of a randomly chosen mode $f\in{\cal N}(\mathbb{R}^{2m})$ and the 15 complementary modes. More notably, we observe that the purity $\mu$ of the reduced photon-added and -subtracted states is always lower than the Gaussian state's purity $\mu_0$. This is compelling numerical evidence that, for any squeezed vacuum and any given mode basis, entanglement never decreases through the addition or subtraction of a photon.

According to Section \ref{sec:EntanglementAddSub}, we must observe entanglement in {\em every} mode basis, provided the mode in which the photon is subtracted or added is not an eigenmode of the squeezed vacuum, i.e.~$g$ is not an eigenvector of $V_s$. Fig.~\ref{fig:ExpEnt1} indicates that photon subtraction and addition increase entanglement as compared to the initial Gaussian state. This implies that any bipartition of modes which is entangled for the initial Gaussian state will remain entangled after the addition or subtraction process. Hence, it remains to verify the presence of entanglement in the basis eigenmodes of the squeezed vacuum where the initial Gaussian state is fully separable (this basis is unique because we consider a pure state with non-degenerate squeezing). This scenario is considered in Fig.~\ref{fig:ExpEnt2}, where the photon-subtracted and -added squeezed vacuum are reduced to each of the different eigenmodes of $V_s$, sorted according to their respective squeezing. It is seen that photon addition and subtraction entangle a bi-partition of a significantly squeezed significantly squeezed mode and the complementary modes. We note, moreover, that photon subtraction generates more entanglement than photon addition whenever the squeezing is sufficiently high (which is consistent with other recent studies \cite{PhysRevA.93.052313}). Modes with very low squeezing can essentially be interpreted as the vacuum, which limits the effectiveness of photon subtraction. 

It must be emphasised that all modes in Fig.~\ref{fig:ExpEnt2} are part of the same mode basis and that the mode in which the photon is added or subtracted is fixed. Therefore, Fig.~\ref{fig:ExpEnt2} highlights that the state is not fully separable in the basis of eigenmodes of $V_s$, i.e.~its Wigner function cannot be written as (\ref{eq:entGen}). The results in Fig.~\ref{fig:ExpEnt1} indicate the presence of entanglement in any other mode basis. Hence, we have failed to find any basis in which the photon-added or -subtracted state is fully separable. This is consistent with the state being inherently entangled, as predicted in Section \ref{sec:EntanglementAddSub}.

\begin{figure}
\centering
\includegraphics[width=0.39\textwidth]{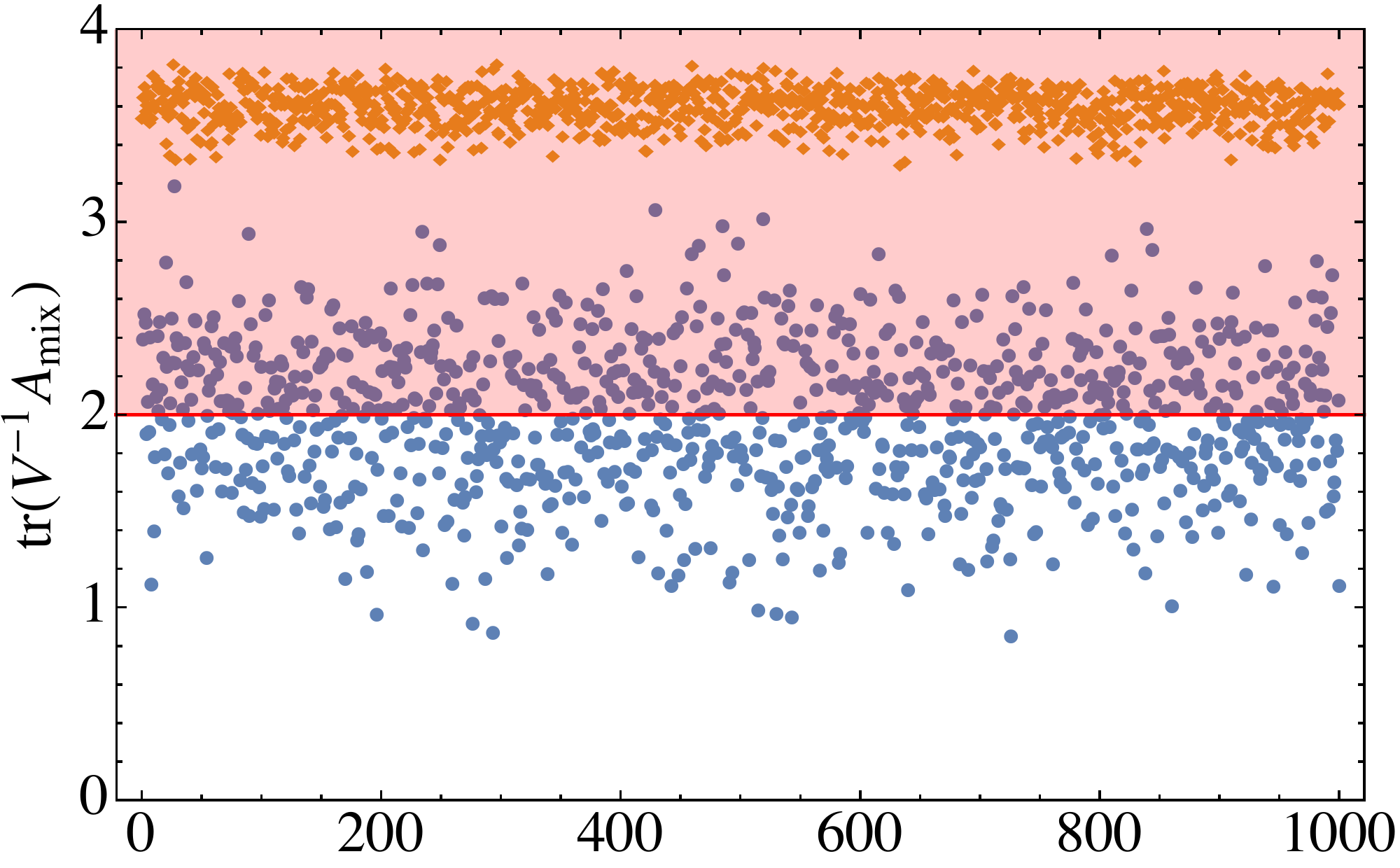}
\includegraphics[width=0.39\textwidth]{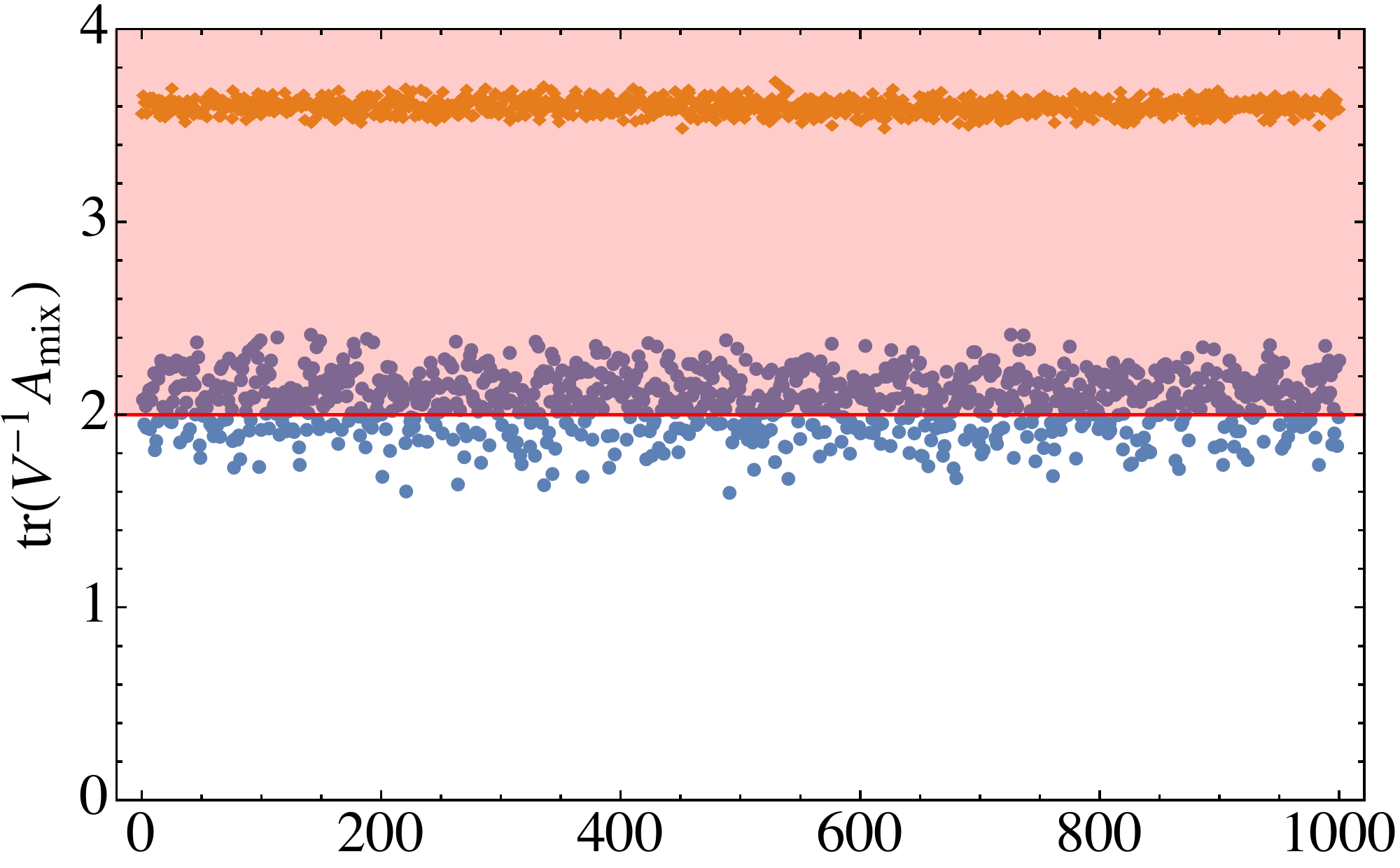}
\includegraphics[width=0.39\textwidth]{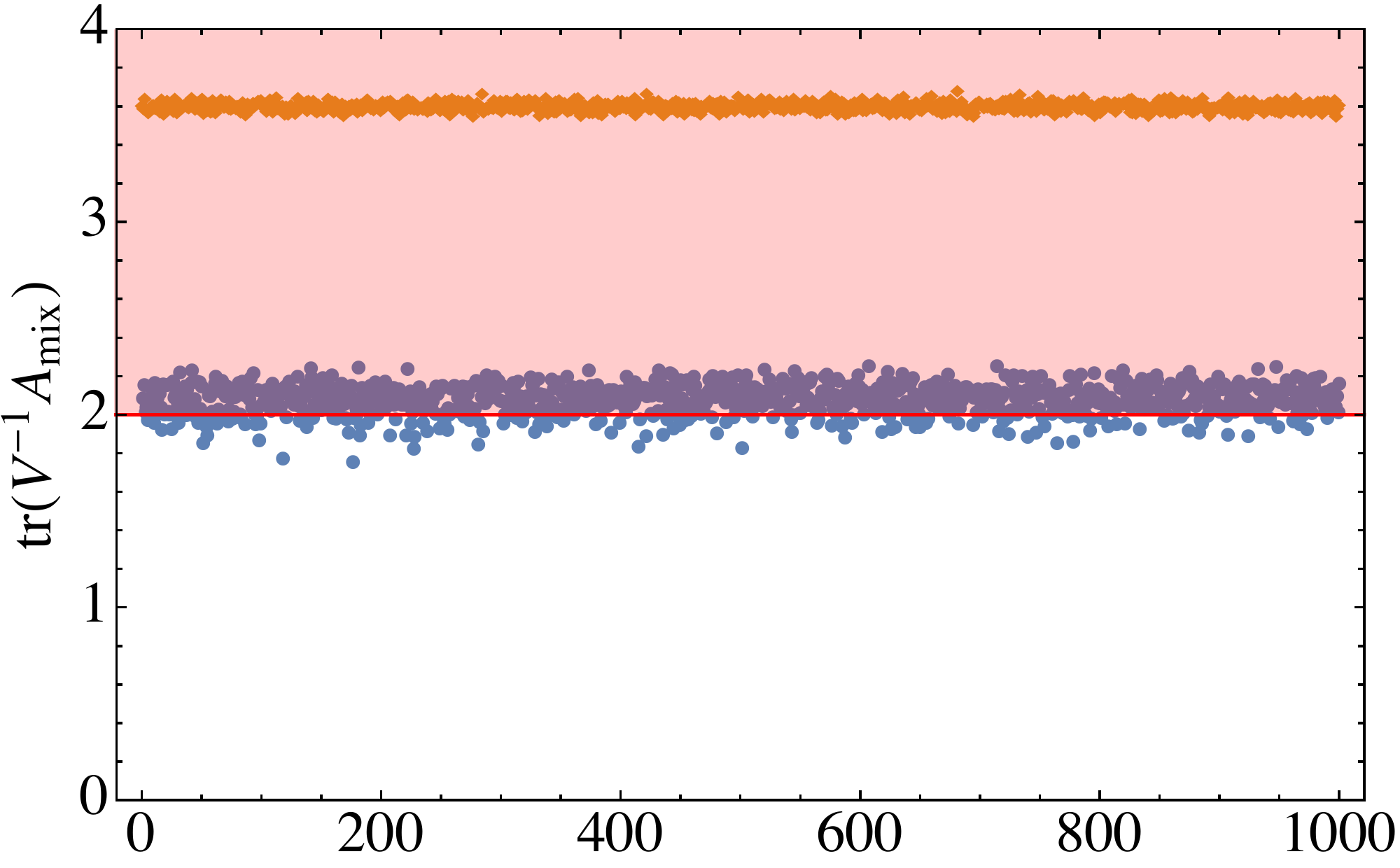}
\includegraphics[width=0.39\textwidth]{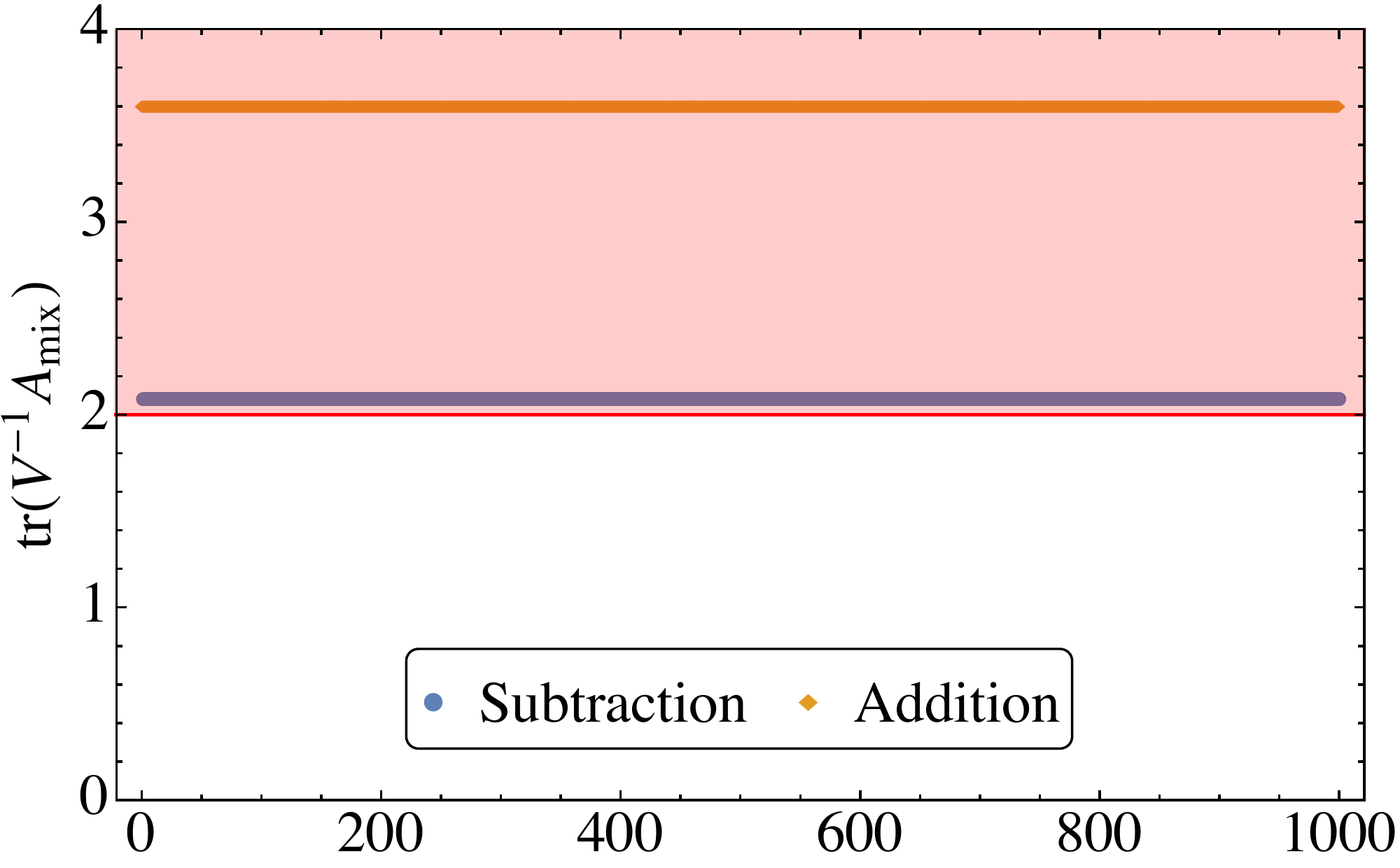}
\caption{Test of negativity condition (\ref{eq:conditionNegat}) for an experimentally obtained state \cite{cai_reconfigurable_2016}, with photon subtraction or addition in a random mode (top) or mixture of random orthogonal modes (others). For every realisation (numbered on horizontal axis) we randomly choose one (top), five (second from top), ten (second from bottom), or sixteen (bottom) orthogonal modes $g_k$ in the mixtures (\ref{eq:photonSubRho}, \ref{eq:photonAddRho}). Weights $\gamma_k$ in the mixtures (\ref{eq:photonSubRho}, \ref{eq:photonAddRho}) are the same for every mode. Only realisations falling in the red zone lead to negative Wigner functions. \label{fig:ExpNeg1}}
\end{figure}

\paragraph*{Negativity---} The negativity of the Wigner function obtained through the {\em pure} addition or subtraction of a photon to the Gaussian state of \cite{cai_reconfigurable_2016} was already studied in \cite{walschaers_entanglement_2017}. Here we treat the case where the addition and subtraction processes are impure, with $A^{\pm}_{\rm mix}$ given by (\ref{eq:AmixMat}). In Fig.~\ref{fig:ExpNeg1}, we probe the negativity witness $\tr(V^{-1}A^{\pm}_{\rm mix})$ for a varying degree of impurity in the addition and subtraction processes. The top panel shows the case for pure subtraction as a reference. Every datapoint corresponds to one randomly generated mode $g \in \mathcal{N}(\mathbb{R}^{32})$, in which the photon is subtracted or added, for which $\tr(V^{-1}A^{\pm}_{\rm mix})$ was evaluated.

Descending through the panels of Fig.~\ref{fig:ExpNeg1} the impurity of photon addition and subtraction is increased. For the second and third plot from the top, five and ten random orthogonal modes, respectively, participate in the process. In concreto, this implies the choice of a set $\{g_1, \dots, g_{10}\}$ random orthogonal vectors in $\mathcal{N}(\mathbb{R}^{32})$. Following (\ref{eq:photonSubRhoMix}) and (\ref{eq:photonAddRhoMix}), each of these modes $g_k$ comes with an associated weight $\gamma_k$, which physically quantifies the probability that a subtracted photon originated from the associated mode (or that the photon is added to the associated mode in the case of addition). For the sake of simplicity, we choose these weights to be uniform over the modes. Thus, we set $\gamma_k = 1/5$ for the mixture of five modes (second panel from the top in Fig.~\ref{fig:ExpNeg1}), and $\gamma_k = 1/10$ in the mixture of ten modes (third panel from the top in Fig.~\ref{fig:ExpNeg1}). 

We observe that the datapoints are less scattered for increasing amounts of impurities. This should not come as a surprise, because increasing impurity implies an averaging over the randomly chosen modes. This leads to the expectation that for sixteen random orthogonal modes, all realisations should coincide, which is confirmed in the bottom panel of Fig.~\ref{fig:ExpNeg1}. In this case, for every datapoint, a photon is subtracted from or added to a balanced mixture of all modes, such that this scenario describes a fully mode-independent photon addition or subtraction. Physically this case is particularly relevant as for photon subtraction it corresponds to the use of a beamsplitter on a set of co-propagating modes. 

In this case, we find that $A^{\pm}_{\rm mix}$ is independent of the the choice of basis in which the mixture $\{g_k\}$ is represented. Hence, none of the properties of the photon-added or -subtracted state depends  From (\ref{eq:AmixMat}), we directly obtain that fully mixed subtraction or addition lead to
\begin{equation}
\tr(V^{-1}A^{\pm}_{\rm mix}) = 2 \left(1 + \frac{\tr (V^{-1} \pm \mathbb{1})}{\tr (V \pm \mathbb{1})}\right).
\end{equation}
Furthermore, in the fully mode-independent scenario, the negativity condition (\ref{eq:conditionNegat}) reduces to \begin{equation}
\label{eq:simpleCheck}
\tr (V^{-1} \pm \mathbb{1}) > 0.
\end{equation}
In the case of $V$ obtained from \cite{cai_reconfigurable_2016}, we obtain that $\tr (V^{-1} - \mathbb{1}) = 0.672702$ for photon subtraction, which clearly satisfies condition (\ref{eq:simpleCheck}). It is remarkable that a highly multimode impure experimentally generated state can still lead to a negative Wigner function upon photon subtraction with a simple beamsplitter setup. We stress, however, that our result refers to the negativity of the full multimode Wigner function, which is difficult to observe in single-mode measurements. Moreover, Fig.~\ref{fig:ExpNeg1} shows that $\tr(V^{-1}A^{\pm}_{\rm mix})$ is only slightly larger than two. This suggests that the region in phase space where the Wigner function becomes negative is small compared to that of a pure photon subtracted state.\\

Hence, we showed that experimentally achieved Gaussian states can lead to negative Wigner functions upon photon subtraction. In the case of photon addition, negativity of the Wigner function is for granted as implied by (\ref{eq:thiscoolone}). Furthermore, we showed that the entanglement properties of the pure squeezed vacuum, extracted from the Gaussian state \cite{cai_reconfigurable_2016} through the Williamson and Bloch-Messiah decompositions, are in agreement with the results in Section \ref{sec:EntanglementAddSub}. It remains an open question whether these entanglement properties persist and can be used when we consider the actual mixed state which is obtained in the experiment.

\section{Conclusions}

\paragraph*{Summary---} We started the analysis of non-displaced multimode photon-added and -subtracted states (where the addition or subtraction is not necessarily pure) by deriving their truncated correlations functions (\ref{eq:TruncFinalMix}) of arbitrary order. The truncated correlations as such suffice to characterise any quantum state. For Gaussian states, in particular, these truncated correlations vanish beyond second order. Hence, we can interpret all truncated correlations beyond second order as clear signatures of the non-Gaussian properties of the state.

However, in this work the truncated correlations were primarily used as a tool to derive the full-state Wigner function for multimode photon-added and -subtracted states (\ref{eq:WignerFinalElegant}). This novel result is also highlighted in \cite{walschaers_entanglement_2017}, and provides a compact and insightful description of a non-displaced multimode photon-added and -subtracted state with arbitrarily many modes. The more general --and more cumbersome-- result for subtraction and addition from a displaced Gaussian state is given in Appendix \ref{eq:SubAddDisp}. 

The non-Gaussian properties of the photon-added and -subtracted states are all encrypted in the polynomial part of the Wigner function (\ref{eq:WignerFinalElegant}). In particular, we obtained an elegant and simple condition (\ref{eq:conditionNegat}) for having a negative Wigner function. Notably, this condition can be used as a tool for selecting the mode in which to subtract a photon. For photon addition, on the other hand, we formally proved that this condition is always fulfilled, such that the Wigner function is always negative. This negativity condition was studied for the concrete examples of a two-mode symmetrically squeezed vacuum and an experimentally obtained Gaussian state in Figs.~\ref{fig:TwoModeNeg} and \ref{fig:ExpNeg1}, respectively.

Finally, we devoted a considerable part of this work to the study of the entanglement properties which can be extracted from the Wigner function (\ref{eq:WignerFinalElegant}). In Section \ref{sec:EntanglementAddSub}, we formalised that subtracting (or adding) a photon in a mode which is part of a mode basis in which the initial Gaussian state is separable will leave the photon-subtracted (or -added) state separable. For pure states, we showed that subtraction (or addition) of the photon in any other mode will induce entanglement. Importantly, this entanglement cannot be undone by passive linear optics operations. In contrast, we stress in Section \ref{sec:Entanglement} that any entanglement in a Gaussian state can always be undone by changing the mode basis through a passive linear optics operation. Furthermore, we evaluated the reduction of the Wigner function to a subset of modes (\ref{eq:redWigner}). For a global pure state, the purity of this reduced state can then be used as a quantitative probe for the entanglement between the mode to which the system is reduced and the modes which were integrated out. These pure state entanglement properties were evaluated in Figs.~\ref{fig:TwoModeEnt}, and \ref{fig:ExpEnt1} and \ref{fig:ExpEnt2}, for a two-mode symmetrically squeezed vacuum and a sixteen-mode squeezed vacuum which is compatible with the experimentally obtained Gaussian state of \cite{cai_reconfigurable_2016}, respectively.

\paragraph*{Outlook---} We started our introduction by emphasising the importance of non-Gaussian states for quantum computation. In this work we have developed a toolbox which is ready to approach concrete quantum information problems and quantum optics experiments. Notably, one may use these techniques for a detailed analysis of single-photon subtraction from a CV cluster state, as used in measurement-based quantum computation \cite{gu_quantum_2009}.

The presented results also impose several new open questions for future research. First, there are still properties of the general Wigner function (\ref{eq:WignerFinalElegant}) which are to be unveiled. Most notably, we think about the negativity volume of the Wigner function, i.e.~the integral of the negative part. We conjecture that the quantity $\tr(V^{-1}A^{\pm}_{\rm mix})$ in (\ref{eq:conditionNegat}) will be proportional to the negativity volume, but this remains to be proven. Moreover, it remains to be understood how a photon subtraction or addition in a particular mode locally affects different modes as represented by the reduction of the Wigner function to these modes (\ref{eq:redWigner}). 

A second open question is the generalisation of our entanglement results to mixed states. As discussed in Section \ref{sec:EntanglementAddSub}, this is not expected to be a straightforward task. A potential route may be to derive bounds on the entanglement in the spirit of \cite{PhysRevLett.96.080502}. 

Finally, it remains a open question how our present results generalise to multi-photon addition and subtraction. A priori, the methods applied here still apply in a more general scenario, but the derivation of the truncated correlations is expected to become a formidable task. Nevertheless, these truncated correlations are a crucial element of our study, because they are directly measurable in state-of-the-art experiments \cite{cai_reconfigurable_2016}.  

\begin{acknowledgements}
This work is supported by the French National Research Agency projects COMB and SPOCQ, and the European Union Grant QCUMbER (no. 665148). C.F. and N.T. are members of the Institut Universitaire de France.
\end{acknowledgements}

\bibliography{Paper_NonGauss.bib}

\clearpage
\widetext
\appendix

\section{Derivation truncated correlation functions}\label{sec:ProofCor}

We prove Eq.~(\ref{eq:TruncFinalMix}) by induction on $k$, which implies that we assume that all even-order truncated correlations up to order $2k-2$ are indeed given by (\ref{eq:TruncFinalMix}). Moreover, we use that for $k=1$ we have the addition expression (\ref{eq:BoomAg}), which was derived explicitly, and that all odd-order truncated correlations vanish.

We start by explicitly writing that
\begin{equation}\label{eq:beginning}
\<Q(f_1)\dots Q(f_{2k})\>_T = \tr\{\rho Q(f_1)\dots Q(f_{2k}) \} - \sum_{p \in {\cal P}^{(2,4,\dots, 2k-2)} }\prod_{i \in p} \<Q(f_{i_1})\dots Q(f_{i_{r}})\>_T,
\end{equation}
where we use the notation ${\cal P}^{(2,4,\dots, 2k-2)}$ to indicate the set of all partitions of the index set $\{1, \dots 2k\}$ where the allowed number of elements in the subsets (that constitute the partitions) is $2, 4, \dots, 2k-2$ (all even orders up to $2k-2$). This implies that we for a partition $p \in {\cal P}^{(2,4,\dots, 2k-2)}$, we cannot fix the number over element in a subset $i \in p$. Therefore, we denote the this number of elements as ``$r$'', where we know that $r$ is {\em even} and smaller than or equal to $2k-2$. Therefore, we know that
\begin{equation}\label{eq:pair}
 \<Q(f_{i_1})\dots Q(f_{i_{r}})\>_T = \delta_{2,r}  \<Q(f_{i_1})Q(f_{i_2})\>_G +  (-1)^{r/2-1} (r/2-1)! \sum_{p' \in {\cal P}_i^{(2)}}\prod_{i' \in p'} (f_{i'_1}, A^{\pm}_{\rm mix} f_{i'_2}),
\end{equation}
with $\delta_{2,r}$ the kronecker-delta. Moreover, we are now considering ${\cal P}_i^{(2)}$ as the set of partitions of the index set $\{i_1, \dots i_r\}$.

Furthermore, using the factorisation properties of a Gaussian state, we can express the term 
\begin{equation}\begin{split}\label{eq:ThisEquationInApp}
\tr\{\rho Q(f_1)\dots Q(f_{2k}) \}
&=\<Q(f_1)\dots Q(f_{2k})\>_G\\ &\qquad+ \sum_{p \in {\cal P}^{(2)}}\sum_{i \in p} \Bigg( (f_{i_1}, A^{\pm}_{\rm mix}f_{i_2})\prod_{j \in p \setminus i} \<Q(f_{j_1})Q(f_{j_2})\>_G\Bigg)\\
&=\sum_{p \in {\cal P}^{(2)}}\prod_{i \in p} \<Q(f_{i_1})Q(f_{i_2})\>_G\\ &\qquad+ \sum_{p \in {\cal P}^{(2)}}\sum_{i \in p} \Bigg((f_{i_1}, A^{\pm}_{\rm mix}f_{i_2})\prod_{j \in p \setminus i} \<Q(f_{j_1})Q(f_{j_2})\>_G\Bigg),\end{split}
\end{equation}
from which it is clear that $\<Q(f_1)\dots Q(f_{2k})\>_T$ contains products of up to $k$ copies of $(f_{i_1}, A^{\pm}_{\rm mix}f_{i_2})$ (where the arguments $f_{i_1}$ and $f_{i_2}$ vary).

To determine the expression for this product, we first focus on the terms which have exactly $k$ copies of $(f_{i_1},A^{\pm}_{\rm mix}f_{i_2})$. These terms are all contained within
\begin{equation}\begin{split}\label{eq:thisone}
&- \sum_{p \in {\cal P}^{(2,4,\dots, 2k-2)} }\prod_{i \in p} \<Q(f_{i_1})\dots Q(f_{i_{r}})\>_T \\
&\quad= - \sum_{p \in {\cal P}^{(2,4,\dots, 2k-2)} }\prod_{i \in p} \Bigg(\delta_{2,r}  \<Q(f_{i_1})Q(f_{i_2})\>_G +  (-1)^{r-1} (r-1)! \sum_{p' \in {\cal P}_i^{(2)}}\prod_{i' \in p'}(f_{i'_1}, A^{\pm}_{\rm mix}f_{i'_2})\Bigg).
\end{split}
\end{equation}
Moreover, it can be seen that the terms $\delta_{2,r}  \<Q(f_{i_1})Q(f_{i_2})\>_G $ can be ignored, because terms with at least one factor of the form $\delta_{2,r}  \<Q(f_{i_1})Q(f_{i_2})\>_G $ cannot contain $k$ factors of the $A$-type. Therefore, the following expression exactly sums up all the terms with $k$ factor of the $A$-type:
\begin{equation}
\begin{split}\label{eq:ThisStep}
 &- \sum_{p \in {\cal P}^{(2,4,\dots, 2k-2)} }\prod_{i \in p}  (-1)^{r/2-1} (r/2-1)! \sum_{p' \in {\cal P}_i^{(2)}}\prod_{i' \in p'} (f_{i'_1}, A^{\pm}_{\rm mix}f_{i'_2}).\\
& \sim  \sum_{p \in {\cal P}^{(2)}}\prod_{i \in p} (f_{i_1}, A^{\pm}_{\rm mix}f_{i_2}) .
\end{split}
\end{equation}
The similarity relation is straightforward to see but all of these terms appear multiple times. We must approach this counting problem in a structural way. To do so, we translate our problem of set-partitioning to an equivalent problem of integer-partitioning \footnote{An integer-partition denotes a way of writing one integer as a sum of other integers. For example $(2,4,6)$ would be an integer-partition of $12 = 2+4+6$.}. Specifically, in our derivation, the index set $\{1, \dots, 2k\}$ of which we have considered the set-partitions can be linked to the integer-partitions of the integer $k$ (because all the subsets in our partitions have an even number of elements). %We will now relate integer-partitions to the set-partitions of the index set, by think of the integer-partition in terms of the number of elements in each subset in a set-partition. 
%, it is practical to divide the number by two. 
For example, when we think of the set-partition $\{\{1,2\};\{3,4,5,6\}\}$, of an index set with $k=3$, we can associate it to the integer-partition $(1,2)$ of the integer $3$. However, also other set-partitions are associated with the integer-partition $(1,2)$, e.g., $\{\{3,6\};\{1,2,4,5\}\}$. These integer-partitions thus represent a class of set-partitions. The procedure (\ref{eq:pair}) to evaluate the truncated correlation functions breaks the subsets, e.g., $\{3,6\}$ and $\{1,2,4,5\}$ in the set-partition $\{\{3,6\};\{1,2,4,5\}\}$, up in pair-partitions. It now follows that one single pair-partition $p' \in {\cal P}^{(2)}$ can be obtained several times within the class of a specific integer-partition. For example $\{\{1,2\}; \{3,6\};\{4,5\}\}$ can be obtained both by breaking up $\{\{1,2\};\{3,4,5,6\}\}$, but also by breaking up $\{\{3,6\};\{1,2,4,5\}\}$ in pair partitions. However, in the end, it is the {\em pair-partition} which determines which specific product of $A^{\pm}_{\rm mix}$ matrix elements it obtained. In (\ref{eq:ThisStep}), for our example, this implies that both $\{\{1,2\};\{3,4,5,6\}\}$ and $\{\{3,6\};\{1,2,4,5\}\}$ induce a term proportional to $(f_1,A^{\pm}_{\rm mix}f_2)(f_3,A^{\pm}_{\rm mix}f_6)(f_4,A^{\pm}_{\rm mix}f_5)$. The crucial point of grouping everything in classes of set-partitions, associated with an integer partition, is that the factors $(-1)^{r/2-1} (r/2-1)!$ are the same for any partition within the class. After all the $r$-values indicated the number of elements in the subsets and therefore the $r/2$ {\em are} the integers constituting the set-partitions.

It is instructive to rewrite (\ref{eq:ThisStep}) in terms of integer-partitions:
\begin{equation}
\begin{split}\label{eq:ThisStep2}
 &- \sum_{p \in {\cal P}^{(2,4,\dots, 2k-2)} }\prod_{i \in p}  (-1)^{r/2-1} (r/2-1)! \sum_{p' \in {\cal P}_i^{(2)}}\prod_{i' \in p'} (f_{i'_1}, A^{\pm}_{\rm mix}f_{i'_2}).
 \\&=- \sum_{\substack{(I_1, \dots, I_q) \\ I_1 + \dots + I_q = k\\ I_1, \dots, I_q < k}}\Bigg(\prod_{j=1}^q  (-1)^{I_j-1} (I_j-1)!  \Bigg)\Bigg( \sum_{p \in {\cal P}^{(2 I_1,\dots, 2I_q)} } \prod_{i \in p} \Bigg\{ \sum_{p' \in {\cal P}_i^{(2)}}\prod_{i' \in p'} (f_{i'_1}, A^{\pm}_{\rm mix} f_{i'_2}) \Bigg\}\Bigg),\end{split}
\end{equation}
where $q$ is not fixed and depends on the integer-partition. We already argued that the specific terms of the form $\prod_{i \in p} (f_{i_1}, A^{\pm}_{\rm mix} f_{i_2})$ occurs several times. What remains is to count their multiplicity for a given set-partition.

Let us now focus on one specific integer-partition $(I_1, \dots, I_q)$. To count the multiplicity of a specific product of $A_g$-functions within this class of partitions. This product is associated with a specific pair-partition $p' \in {\cal P}^{(2)}$, and our counting process will consist out of counting in how many partitions of the class $(I_1, \dots, I_q)$ we can embed this specific pair-partition. This problem is equivalent to asking in how many ways we can group $k$ elements in a subsets of $I_1$, $I_2$, \dots, and $I_q$ elements. Here, these $k$ elements are the pairs in the pair-partition. This combinatoric problem is solved using the multinomial coefficient, such that we find $$\frac{k!}{\prod^q_{j=1}{I_q!}}.$$
An important additional ingredient is that this is independent of the specific pair-partition we choose, they all occur with the same multiplicity within the specific integer-partition $(I_1, \dots, I_q)$. This implies that 
\begin{equation}\label{eq:ThisStep3}
\sum_{p \in {\cal P}^{(2 I_1,\dots, 2I_q)} } \prod_{i \in p} \Bigg\{ \sum_{p' \in {\cal P}_i^{(2)}}\prod_{i' \in p'} (f_{i'_1}, A^{\pm}_{\rm mix}f_{i'_2}) \Bigg\} = \frac{k!}{\prod^q_{j=1}{I_q!}}\sum_{p \in {\cal P}^{(2)}}\prod_{i \in p} (f_{i_1}, A^{\pm}_{\rm mix}f_{i_2}).
\end{equation}
If we now insert (\ref{eq:ThisStep3}) in (\ref{eq:ThisStep2}), we find
\begin{equation}
\begin{split}\label{eq:ThisStep4}
 &- \sum_{p \in {\cal P}^{(2,4,\dots, 2k-2)} }\prod_{i \in p}  (-1)^{r/2-1} (r/2-1)! \sum_{p' \in {\cal P}_i^{(2)}}\prod_{i' \in p'} (f_{i'_1}, A^{\pm}_{\rm mix} f_{i'_2}).
 \\&=- \sum_{\substack{(I_1, \dots, I_q) \\ I_1 + \dots + I_q = k\\ I_1, \dots, I_q < k}}\Bigg(\prod_{j=1}^q  (-1)^{I_j-1} (I_j-1)!  \Bigg)\Bigg(\frac{k!}{\prod^q_{j=1}{I_q!}}\sum_{p \in {\cal P}^{(2)}}\prod_{i \in p} (f_{i_1}, A^{\pm}_{\rm mix}f_{i_2})\Bigg)
  \\&=- \Bigg(k! \sum_{\substack{(I_1, \dots, I_q) \\ I_1 + \dots + I_q = k\\ I_1, \dots, I_q < k}}\prod^q_{j=1}\frac{(-1)^{I_j-1}}{{I_q}}\Bigg) \sum_{p \in {\cal P}^{(2)}}\prod_{i \in p} (f_{i_1}, A^{\pm}_{\rm mix}f_{i_2}).\end{split}
\end{equation}
Finally, we use that
\begin{equation}\label{eq:thespecialone}
\sum_{\substack{(I_1, \dots, I_q) \\ I_1 + \dots + I_q = k\\ I_1, \dots, I_q \leqslant k}}\prod^q_{j=1}\frac{(-1)^{I_j-1}}{{I_q}} =0,
\end{equation}
which explicitly includes the integer-partition $(k)$ of $k$, which was excluded in the derivation of (\ref{eq:ThisStep4}). This implies that
\begin{equation}\label{eq:ThisStep5}
\sum_{\substack{(I_1, \dots, I_q) \\ I_1 + \dots + I_q = k\\ I_1, \dots, I_q < k}}\prod^q_{j=1}\frac{(-1)^{I_j-1}}{{I_q}} = - \frac{(-1)^{k-1}}{k}.
\end{equation}
Inserting (\ref{eq:ThisStep5}) in (\ref{eq:ThisStep4}) therefore results in 
\begin{equation}
\begin{split}\label{eq:ThisStep6}
 &- \sum_{p \in {\cal P}^{(2,4,\dots, 2k-2)} }\prod_{i \in p}  (-1)^{r/2-1} (r/2-1)! \sum_{p' \in {\cal P}_i^{(2)}}\prod_{i' \in p'} (f_{i'_1}, A^{\pm}_{\rm mix} f_{i'_2}).
 \\&= (-1)^{k-1} (k-1)! \sum_{p \in {\cal P}^{(2)}}\prod_{i \in p} (f_{i_1}, A^{\pm}_{\rm mix} f_{i_2}),\end{split}
\end{equation}
which is the final result for the contribution of the terms with only factors of the form $(f_{i_1}, A^{\pm}_{\rm mix}f_{i_2})$.\\

It now remains to show that all other terms, in which factors of the form $\<Q(f_{i_1})Q(f_{i_2}))\>_G$ appear, vanish. We will use primarily the same apparatus to prove this, and start by observing that only integer partitions with a component which is $1$ can lead to a term $\<Q(f_{i_1})Q(f_{i_2}))\>_G$. Therefore, we focus on a class associated with the integer-partition $$(\underbrace{1,\dots, 1 }_{x < k-1}, I_1, \dots I_{q'}), \quad \text{with} \quad I_1 + \dots + I_{q'} = k - x,$$
where the $x$ first terms are particularly associated with a factor $\<Q(f_{i_1})Q(f_{i_2}))\>_G$. We exclude the cases $x = k$ and $x = k-1$ because there $\tr\{\rho Q(f_1)\dots Q(f_{2k}) \}$ in (\ref{eq:beginning}) must also be considered. We can group the terms with $x$ factors $\<Q(f_{i_1})Q(f_{i_2}))\>_G$ in (\ref{eq:thisone}) as
\begin{equation}\begin{split}
- &\sum_{\substack{(1, \dots, 1, I_1, \dots, I_{q'}) \\ I_1 + \dots + I_{q,} = k-x\\ I_1, \dots, I_{q'} \leqslant k-x }} \sum_{\substack{ {\cal X} \subset {\cal I}\\ \# {\cal X} = 2x}} \Bigg(  \sum_{p \in {\cal P}_{\cal X}^{(2)}}\prod_{i \in p} \<Q(f_{i_1})Q(f_{i_2})\>_G \Bigg)
\\ &\qquad\qquad\times \Bigg(\prod_{j=1}^{q'}  (-1)^{I_j-1} (I_j-1)!  \Bigg) \Bigg( \sum_{p \in {\cal P}_{{\cal I} \setminus {\cal X} }^{(2 I_1,\dots, 2I_{q'})} } \prod_{i \in p} \Bigg\{ \sum_{p' \in {\cal P}_i^{(2)}}\prod_{i' \in p'} (f_{i'_1},A^{\pm}_{\rm mix} f_{i'_2}) \Bigg\}\Bigg),\\
&=- \sum_{\substack{ {\cal X} \subset {\cal I}\\ \# {\cal X} = 2x}} \Bigg(  \sum_{p \in {\cal P}_{\cal X}^{(2)}}\prod_{i \in p} \<Q(f_{i_1})Q(f_{i_2})\>_G \Bigg)
\\ &\qquad\times \sum_{\substack{(I_1, \dots, I_{q'}) \\ I_1 + \dots + I_{q,} = k-x\\ I_1, \dots, I_{q'} \leqslant k-x }} \Bigg(\prod_{j=1}^{q'}  (-1)^{I_j-1} (I_j-1)!  \Bigg) \Bigg( \sum_{p \in {\cal P}_{{\cal I} \setminus {\cal X} }^{(2 I_1,\dots, 2I_{q'})} } \prod_{i \in p} \Bigg\{ \sum_{p' \in {\cal P}_i^{(2)}}\prod_{i' \in p'} (f_{i'_1}, A^{\pm}_{\rm mix} f_{i'_2}) \Bigg\}\Bigg),
\end{split}
\end{equation}
where we introduce ${\cal I} = \{1, 2, \dots, 2k \}$. The set ${\cal X}$ contains the subset of indices with which we associate the factors $\<Q(f_{i_1})Q(f_{i_2})\>_G$, the other indices ${\cal I} \setminus {\cal X}$ have a factor of the $A$-type connected to them. We introduce the notation ${\cal P}^{(2)}_{\cal X}$ to denote the pair-partitions of the set ${\cal X}$, whereas ${\cal P}_{{\cal I} \setminus {\cal X} }^{(2 I_1,\dots, 2I_{q'})} $ are all the possible even partitions of ${\cal I} \setminus {\cal X} $.

We can limit our efforts to understanding that
\begin{equation}
\begin{split}\label{eq:complicatedZero}
&\sum_{\substack{(I_1, \dots, I_{q'}) \\ I_1 + \dots + I_{q,} = k-x\\ I_1, \dots, I_{q'} \leqslant k-x }}\Bigg(\prod_{j=1}^{q'}  (-1)^{I_j-1} (I_j-1)!  \Bigg) \Bigg( \sum_{p \in {\cal P}_{{\cal I} \setminus {\cal X} }^{(2 I_1,\dots, 2I_{q'})} } \prod_{i \in p} \Bigg\{ \sum_{p' \in {\cal P}_i^{(2)}}\prod_{i' \in p'} (f_{i'_1}, A^{\pm}_{\rm mix} f_{i'_2}) \Bigg\}\Bigg)\\
&\quad= \sum_{\substack{(I_1, \dots, I_{q'}) \\ I_1 + \dots + I_{q,} = k-x\\ I_1, \dots, I_{q'} \leqslant k-x }}\Bigg(\prod_{j=1}^{q'}  (-1)^{I_j-1} (I_j-1)!  \Bigg) \Bigg( \frac{(k-x)!}{\prod_{j=1}^{q'} I_j!} \sum_{p \in {\cal P}_{{\cal I} \setminus {\cal X} }^{(2)} } \prod_{i \in p} (f_{i_1}, A^{\pm}_{\rm mix} f_{i_2}) \Bigg)\\
&\quad= \Bigg(\sum_{\substack{(I_1, \dots, I_{q'}) \\ I_1 + \dots + I_{q,} = k-x\\ I_1, \dots, I_{q'} \leqslant k-x }}\prod_{j=1}^{q'}  (-1)^{I_j-1}   \frac{(k-x)!}{I_j} \Bigg)\sum_{p \in {\cal P}_{{\cal I} \setminus {\cal X} }^{(2)} } \prod_{i \in p} (f_{i_1}, A^{\pm}_{\rm mix} f_{i_2})\\
&\quad=0.
\end{split}
\end{equation}
The above steps use exactly the same reasoning as the derivation of (\ref{eq:ThisStep4}). Finally, we used (\ref{eq:thespecialone}) to obtain that the expression is zero.

The case where $x=k$ can be treated explicitly. This contribution reads
\begin{equation}\label{eq:GaussianTerm}
\<Q(f_1)\dots Q(f_{2k})\>_G -   \sum_{p \in {\cal P}^{(2)}}\prod_{i \in p} \<Q(f_{i_1})Q(f_{i_2})\>_G = 0,
\end{equation}
because it is the truncated correlation function in a Gaussian state. The case $x = k-1$, i.e., where there is exactly one factor of the form $A_g$, leads to a contribution
\begin{equation}
\begin{split} \label{eq:lastterm}
&\sum_{p \in {\cal P}^{(2)}}\sum_{i \in p} \Bigg( (f_{i_1}, A^{\pm}_{\rm mix} f_{i_2})\prod_{j \in p \setminus i} \<Q(f_{j_1})Q(f_{j_2})\>_G\Bigg) \\
&-\sum_{\substack{(1, \dots, 1, I_1, \dots, I_{q'}) \\ I_1 + \dots + I_{q,} = k-x\\ I_1, \dots, I_{q'} \leqslant k-x }} \sum_{\substack{ {\cal X} \subset {\cal I}\\ \# {\cal X} = 2k-2}} \Bigg(  \sum_{p \in {\cal P}_{\cal X}^{(2)}}\prod_{i \in p} \<Q(f_{i_1})Q(f_{i_2})\>_G \Bigg)
\\ &\qquad\qquad\times\Bigg( \sum_{p \in {\cal P}_{{\cal I} \setminus {\cal X} }^{(2 I_1,\dots, 2I_{q'})} } \prod_{i \in p} \Bigg\{ \sum_{p' \in {\cal P}_i^{(2)}}\prod_{i' \in p'} (f_{i'_1}, A^{\pm}_{\rm mix} f_{i'_2}) \Bigg\}\Bigg)\\
&= 0.
\end{split}
\end{equation}
This contribution vanishes because in this case ${\cal I} \setminus {\cal X} = \{i_1, i_2\}$. Therefore, the second sum is equivalent to the first. \\

We have now evaluated all the different terms in (\ref{eq:beginning}), assuming that for lower orders (\ref{eq:pair}) holds. Indeed, for our final result, we have
\begin{equation}\begin{split}
\<Q(f_1)\dots Q(f_{2k})\>_T &= (\ref{eq:ThisStep6}) + (\ref{eq:complicatedZero}) + (\ref{eq:GaussianTerm}) + (\ref{eq:lastterm})\\
&=  (-1)^{k-1} (k-1)! \sum_{p \in {\cal P}^{(2)}}\prod_{i \in p} (f_{i_1}, A^{\pm}_{\rm mix} f_{i_2}) \quad \text{for $k > 1$}.
\end{split}
\end{equation}
This concludes the derivation of (\ref{eq:TruncFinalMix}).
\newpage
\section{Computation of the Wigner function (\ref{eq:WignerFinalElegant})}\label{app:B}

We treat the problem explicitly in the basis of eigenvectors ${\cal E} = \{e_1, \dots, e_{2m}\}$ of $V$. Note that these eigenvectors in general do not respect the symplectic structure of phase space. The following steps are merely technical tricks to evaluate the Fourier transform, and cannot be directly connected to a well-defined mode-space.
%In the simplest scenario, we subtract a photon from one of these supermodes. Formally, this means that there exists an $i$ for which $g \in {\rm span}(e^{(i)}_x,e^{(i)}_p)$. As in the single mode case, all the different superpositions of $e^{(i)}_x$ and $e^{(i)}_p$ that can be chosen for $g$ lead to the same result; what is relevant is the mode $i$. From (\ref{eq:AgMat}) it now follows that, both,
%\begin{equation}\label{eq:orthoAg}
%A_g(e_q^{(i)},e_{q'}^{(j)}) = 0 \, \text{ for $i \neq j$ or $q \neq q'$},
%\end{equation}
%and
%\begin{equation}\label{eq:hereOtherOrtho}
%A_{e^{(i)}_{x/p}}(e_q^{(j)},e_q^{(j)})=0, \quad \text{if $i\neq j$}.
%\end{equation}
In (\ref{eq:wignerGen}), we may write 
\begin{equation}\label{eq:BasisExpansion}\alpha = \sum_{j=1}^{2m} \alpha_j e_j\quad \text{and}\quad \beta=\sum_{j=1}^{2m}  \beta_j e_j.\end{equation}
%The presence of cross-terms in (\ref{eq:wignerGen}), or equivalently the impossibility to jointly diagonalise $V$ and $A_g$, in general forbids us to factorise the Wigner function (\ref{eq:Factorisation}). To show this, we formally write the Wigner function. 
We start from (\ref{eq:BasisExpansion}), where it now follows that
\begin{equation}\label{eq:againAgalphalpha}
(\alpha,A^{\pm}_{\rm mix}\alpha) = \sum_{i,j=1}^m \alpha_i \alpha_j (e_i, A^{\pm}_{\rm mix}e_j)
\end{equation}
Upon inserting (\ref{eq:againAgalphalpha})  in (\ref{eq:wignerGen}) and using the linearity of the integration we find explicitly
\begin{equation}
\begin{split}
W(\beta_1, \dots, \beta_{2m}) =&  \prod_{k=1}^{2m}\frac{1}{2\pi}\int_{-\infty}^{\infty}{\rm d}\alpha_k\,  \exp\left\{-\frac{v_k {\big(\alpha_k\big)}^2}{2} - i\alpha_k \beta_k \right\}\\
&-\sum_{j=1}^{2m} \frac{(e_j,A^{\pm}_{\rm mix}e_j)}{8\pi^2} \int_{-\infty}^{\infty}{\rm d}\alpha_j\, \big(\alpha_j\big)^2 \exp\left\{-\frac{v_j {\big(\alpha_j\big)}^2}{2} - i\alpha_j \beta_j\right\}\\
& \qquad \times \prod_{\substack{k=1\\k \neq j}}^{2m}\frac{1}{2\pi}\int_{-\infty}^{\infty}{\rm d}\alpha^{(j)}_q\,  \exp\left\{-\frac{v^{(j)}_q {\big(\alpha^{(j)}_q\big)}^2}{2} - i\alpha^{(j)}_q \beta^{(j)}_q \right\}\\
&-\sum_{\substack{j,j'=1 \\ j \neq j'}}^{2m} \frac{(e_j, A^{\pm}_{\rm mix}e_{j'})}{8\pi^2} \int_{-\infty}^{\infty}{\rm d}\alpha_j{\rm d}\alpha_{j'}\, \alpha_{j}\alpha_{j'} \exp\left\{-\frac{v_j {\big(\alpha_j\big)}^2}{2}-\frac{v_{j'} {\big(\alpha_{j'}\big)}^2}{2} - i\alpha_j \beta_j - i\alpha_{j'} \beta_{j'} \right\}\\
& \qquad \times \prod_{\substack{k=1\\ k \neq j,\, k \neq j'}}^{2m}\frac{1}{2\pi}\int_{-\infty}^{\infty}{\rm d}\alpha_k\,  \exp\left\{-\frac{v_k {\big(\alpha_k\big)}^2}{2} - i\alpha_k \beta_k \right\}\\
=& \frac{1}{(2\pi)^m}\Bigg(1 + \sum_{j=1}^m \frac{(e_j, A^{\pm}_{\rm mix} e_j) \Big[\big(\beta_j\big)^2 - v_j\Big]}{2\big(v_j\big)^{2}} + \sum_{\substack{j,j'=1\\j \neq j'}}^{2m}\frac{(e_j,A^{\pm}_{\rm mix} e_{j'}) \beta_j\beta_{j'}}{2v_j v_{j'}} \Bigg)\prod_{j=1}^{2m}\frac{1}{\sqrt{v_j}}\exp\Big\{-\frac{\big(\beta_j\big)^2}{2 v_j}\Big\}.
\end{split}
\end{equation}
The last step consists simply out of evaluating the Fourier transforms and grouping the terms. It must be stressed that, since this basis is not symplectic, we cannot interpret the Wigner function in this form as a quasi-probability distribution on phase space. Therefore we require a basis-independent way of representing the function. This can be obtained by regrouping the terms and observing that 
\begin{equation}
\begin{split}
\sum_{j=1}^m \frac{(e_j, A^{\pm}_{\rm mix} e_j)\big(\beta_j\big)^2}{2\big(v_j\big)^{2}} + \sum_{\substack{j,j'=1\\j \neq j'}}^{2m}\frac{(e_j,A^{\pm}_{\rm mix}e_{j'}) \beta_j\beta_{j'}}{2v_j v_{j'}} &= \sum_{\substack{j,j'=1}}^{2m}\frac{(e_j,A^{\pm}_{\rm mix}e_{j'}) \beta_j\beta_{j'}}{2v_j v_{j'}}
\\&=\frac{1}{2}\sum_{\substack{j,j'=1}}^{2m} \beta_j\beta_{j'} (e_j, V^{-1}A^{\pm}_{\rm mix}V^{-1} e_{j'})
\\&=\frac{1}{2} \left( \sum_j \beta_j e_j, V^{-1}A^{\pm}_{\rm mix}V^{-1} \sum_{j'} \beta_{j'} e_{j'}  \right)
\\&= \frac{1}{2} \left(\beta, V^{-1}A^{\pm}_{\rm mix}V^{-1} \beta  \right),
\end{split}
\end{equation}
where we use the linearity of the inner product and that $V e_j = v_j e_j$. Similarly, we obtain 
\begin{align}
&-\sum_{j=1}^m\frac{\big(\beta_j\big)^2}{2 v_j} = -\frac{(\beta,V^{-1} \beta)}{2}\\
&- \sum_{j=1}^m \frac{(e_j, A^{\pm}_{\rm mix} e_j) }{2 v_j} = \frac{1}{2} \tr\{V^{-1} A^{\pm}_{\rm mix}\},
\end{align}
such that ultimately 
\begin{equation}\label{eq:WignerFinalElegantApp}
W(\beta) = \frac{1}{2^{m+1} \pi^m \sqrt{\det V}} \Bigg((\beta, V^{-1} A^{\pm}_{\rm mix} V^{-1} \beta) -  \tr(V^{-1}A^{\pm}_{\rm mix}) + 2 \Bigg)e^{-\frac{1}{2}(\beta, V^{-1} \beta)}.
\end{equation}

\newpage
\section{Photon-addition and subtraction with displaced states}\label{eq:SubAddDisp}

The discussion in Section \ref{sec:EntanglementAddSub} requires an expression for the Wigner function for a displaced squeezed vacuum from/to which a photon is subtracted/added. In this Appendix, we go one step beyond this need and we derive the Wigner function for photon subtraction or addition in a general displaced Gaussian state which need not be pure. Our derivation exploits the previously obtained result (\ref{eq:WignerFinalElegant}) for the non-displaced case.\\

Any displaced Gaussian state can be written as 
\begin{equation}
\rho_{\xi} = D(\xi)\rho_G D(-\xi),
\end{equation}
where $\rho_G$ is a non-displaced Gaussian state, characterised by a covariance matrix $V$. A photon-subtracted state can then be written as 
\begin{equation}
\rho = \frac{a(g)D(\xi)\rho_G D(-\xi)a^{\dag}(g)}{\<D(-\xi)\hat{n}(g)D(\xi)\>_G},
\end{equation}
with $\hat{n}(g)$ the number operators in mode $g$. It is not hard to evaluate that \begin{equation}\<D(-\xi)\hat{n}(g)D(\xi)\>_G = \<\hat{n}(g)\>_G + \frac{1}{4}\big((\xi, g)^2 + (\xi, Jg)^2\big).\end{equation} We may now use that \begin{equation}D(\xi)a(g) = a(g)D(\xi) - \frac{1}{2}\big(\xi,(\mathbb{1}+iJ) g\big)\,D(\xi), \quad \text{and}\quad D(\xi)a^{\dag}(g) = a^{\dag}(g)D(\xi) - \frac{1}{2}\big(\xi,(\mathbb{1}-iJ) g\big)\,D(\xi),\end{equation} and  focus on 
\begin{equation}\begin{split}
a(g)D(\xi)\rho_G D(-\xi)a^{\dag}(g) = &D(\xi)a(g)\rho_G a^{\dag}(g) D(-\xi) + \frac{1}{4}\big((\xi,g)^2 + (\xi, Jg)^2\big) D(\xi)\rho_G D(-\xi)\\
&+ \frac{1}{2} \big(\xi,(\mathbb{1}-iJ) g\big) D(\xi)a(g)\rho_G  D(-\xi) + \frac{1}{2}\big(\xi,(\mathbb{1}+iJ) g\big) D(\xi)\rho_G a^{\dag}(g) D(-\xi).
\end{split}
\end{equation}
Next, we use that $D(-\xi)D(2J\!\alpha)D(\xi)= e^{i(\xi,\alpha)}D(2J\!\alpha)$ to write
\begin{equation}\begin{split}
\chi(\alpha) = \tr\{D(2J\alpha) \rho\} = \frac{ e^{i(\xi,\alpha)} }{ \<\hat{n}(g)\>_G + \frac{1}{4}\big((\xi, g)^2 + (\xi, Jg)^2\big)} \Bigg(& \<\hat{n}(g)\>_G \chi_{-}(\alpha) + \frac{1}{4}\big( (\xi, g)^2 + (\xi, Jg)^2 \big)\chi_{G}(\alpha)\\
&+ \frac{1}{2}\big(\xi,(\mathbb{1}-iJ) g\big) \tr\{D(2J\!\alpha)a(g)\rho_G \} \\
&+ \frac{1}{2}\big(\xi,(\mathbb{1}+iJ) g\big) \tr\{a^{\dag}(g)D(2J\!\alpha)\rho_G \}\Bigg)
\end{split}
\end{equation}
Now we use that $\big(\xi,(\mathbb{1}-iJ) g\big)a(g) = a\big((P_g+P_{Jg})\xi\big)$ and $\big(\xi,(\mathbb{1}+iJ) g\big)a^{\dag}(g) = a^{\dag}\big((P_g+P_{Jg})\xi\big)$ and that $D(2J\!\alpha)a\big((P_g+P_{Jg})\xi\big) = a\big((P_g+P_{Jg})\xi\big)D(2J\!\alpha) - \big(\xi,(P_g+P_{Jg})(J+i\mathbb{1})\alpha\big)D(2J\!\alpha)$, such that
\begin{equation}\begin{split}
\big(\xi,(\mathbb{1}-iJ) g\big) \tr\{D&(2J\!\alpha)a(g)\rho_G \} + \big(\xi,(\mathbb{1}+iJ) g\big) \tr\{a^{\dag}(g)D(2J\!\alpha)\rho_G \} \\
&= \tr\{Q\big((P_g+P_{Jg})\xi\big) D(2J\!\alpha) \rho_G\} - \big(\alpha,(-J+i\mathbb{1}) (P_g+P_{Jg})\xi\big)\chi_G(\alpha),
\end{split}
\end{equation}
and
\begin{equation}\begin{split}
\chi(\alpha) = \frac{ e^{i(\xi,\alpha)} }{ \<\hat{n}(g)\>_G + \frac{1}{4}\big((\xi, g)^2 + (\xi, Jg)^2\big)} \Bigg(& \<\hat{n}(g)\>_G \chi_{-}(\alpha) + \bigg( \frac{1}{4}\big((\xi, g)^2 + (\xi, Jg)^2\big) - \frac{1}{2}\big(\xi,(P_g+P_{Jg})(J+i\mathbb{1})\alpha\big) \bigg)\chi_{G}(\alpha)\\
&+ \frac{1}{2}\tr\big\{Q\big((P_g+P_{Jg})\xi\big) D(2J\!\alpha) \rho_G\big\}\Bigg).
\end{split}
\end{equation}
In order to proceed to evaluating the Wigner function, it remains to evaluate $\tr\{Q\big((P_g+P_{Jg})\xi\big) D(2J\!\alpha) \rho_G\}$. We set $x = (P_g+P_{Jg})\xi$ and evaluate
\begin{equation}
\tr\{Q(x) D(2J\!\alpha) \rho_G\} = \tr\{Q(x) \exp\{i Q(\alpha)\} \rho_G\} = \tr\{Q(x)\rho_G\} + i \tr\{Q(x)Q(\alpha)\rho_G\}  -\frac{1}{2}\tr\{Q(x)Q(\alpha)^2\rho_G\} -\frac{i}{6}\tr\{Q(x)Q(\alpha)^3\rho_G\} + \dots
\end{equation}
We observe that, because $\rho_G$ is a non-displaced state, all terms with an odd number of $Q$ operators vanish. This leaves terms proportional to
\begin{equation}
\tr\{Q(x)Q(\alpha)^{2k+1}\rho_G\} = (2k+1)(2k-1)!! \tr\{Q(x)Q(\alpha)\rho_G\}\tr\{Q(\alpha)^2\rho_G\}^k = (2k+1)(2k-1)!! [(x,V\alpha)-i(x,J\alpha)] (\alpha,V\alpha)^k,
\end{equation}
where we use explicitly that $\rho_G$ is a non-displaced Gaussian state with covariance matrix $V$, such that its correlations factorise. This implies that
\begin{equation}\begin{split}
\tr\{Q(x) D(2J\!\alpha) \rho_G\} &= i\sum_{k=0}^{\infty}\frac{(-1)^k}{(2k+1)!}(2k+1)!! [(x,V\alpha)-i(x,J\alpha)](\alpha,V\alpha)^k\\ &=  i\big((x,V\alpha)-i(x,J\alpha)\big) \sum_{k=0}^{\infty}\frac{1}{k!} \left(\frac{-(\alpha,V\alpha)}{2}\right)^k\\&= i\big((x,V\alpha)-i(x,J\alpha)\big) \exp\left\{\frac{-(\alpha,V\alpha)}{2}\right\}.
\end{split}\end{equation}

This ultimately allows us to rewrite
\begin{equation}\begin{split}
\chi(\alpha) = \frac{ e^{i(\xi,\alpha)} }{ \<\hat{n}(g)\>_G + \frac{1}{4}\big((\xi, g)^2 + (\xi, Jg)^2\big)} \Bigg(& \<\hat{n}(g)\>_G \chi_{-}(\alpha) + \frac{1}{4}\big( (\xi, g)^2 + (\xi, Jg)^2 \big) \chi_{G}(\alpha) \\
&+\frac{1}{2}\bigg(\xi, \big( (P_g+P_{Jg})J +i (P_g+P_{Jg})V - (P_{Jg} + P_{g})(J+i\mathbb{1})\big) \alpha \bigg)\\
&\qquad\qquad\times\exp\left\{\frac{-(\alpha,V\alpha)}{2}\right\}\Bigg).
\end{split}
\end{equation}
The Fourier transformation which leads to the Wigner function can be carried out straightforwardly in the basis where $V$ is diagonal.  This leads us to
\begin{equation}\begin{split}
W^-_{\xi}(\beta) =& \frac{1 }{ \<\hat{n}(g)\>_G + \frac{1}{4}\big((\xi, g)^2 + (\xi, Jg)^2\big)} \Bigg( \<\hat{n}(g)\>_G W_{-}(\beta - \xi) + \frac{1}{4}\big( (\xi, g)^2 + (\xi, Jg)^2 \big) W_{G}(\beta - \xi) \\
&\qquad\qquad\qquad\qquad\qquad\qquad+  \frac{1}{2}\big(\xi, (P_g+P_{Jg})(\mathbb{1}-V^{-1}) (\beta-\xi) \big)\frac{\exp\left\{\frac{-(\beta-\xi,V^{-1}\beta-\xi)}{2}\right\}}{(2\pi)^m\sqrt{\det V}}\Bigg)\\
&=  \frac{W_{G}(\beta - \xi) }{ 2\<\hat{n}(g)\>_G + \frac{1}{2}\big((\xi, g)^2 + (\xi, Jg)^2\big)} \Bigg(\<\hat{n}(g)\>_G \Big[([\beta-\xi], V^{-1} A^-_{g} V^{-1} [\beta-\xi]) -  \tr(V^{-1}A^-_{\rm g}) + 2 \Big]\\&
 \qquad\qquad\qquad\qquad\qquad\qquad +\frac{1}{2}\big( (\xi, g)^2 + (\xi, Jg)^2 \big) + \big(\xi, (P_g+P_{Jg})(\mathbb{1}-V^{-1}) (\beta-\xi) \big)  \Bigg),
\end{split}
\end{equation}
which is the final Wigner function for photon-subtraction from a displaced state.

A completely analogous calculation can be performed for photon addition. Here the starting point is
\begin{equation}
\rho = \frac{a^{\dag}(g)D(\xi)\rho_G D(-\xi)a(g)}{\<D(-\xi)\hat{n}(g)D(\xi)\>_G+1},
\end{equation}
going through the same steps of calculation, while taking into account the changes in signs, leads us to
\begin{equation}\begin{split}
\chi(\alpha) = \frac{ e^{i(\xi,\alpha)} }{ \<\hat{n}(g)\>_G+ 1 + \frac{1}{4}\big((\xi, g)^2 + (\xi, J\,g)^2\big)} \Bigg(& (\<\hat{n}(g)\>_G + 1) \chi_{+}(\alpha) + \frac{1}{4}\big( (\xi, g)^2 + (\xi, Jg)^2 \big) \chi_{G}(\alpha) \\
&+\frac{1}{2}\Big(\xi, \big((P_{Jg} + P_{g})(J+i\mathbb{1}) - (P_g+P_{Jg})J +i (P_g+P_{Jg})V\big) \alpha \Big) \chi_{G}(\alpha)\Bigg).
\end{split}
\end{equation}
This directly leads to a Wigner function of the form
\begin{equation}\begin{split}
W^+_{\xi}(\beta) &=  \frac{W_{G}(\beta - \xi)}{ \<\hat{n}(g)\>_G+1 + \frac{1}{4}\big((\xi, g)^2 + (\xi, Jg)^2\big)} \Bigg(\frac{\<\hat{n}(g)\>_G+1}{2} \Big[\big((\beta-\xi), V^{-1} A^+_{g} V^{-1} (\beta-\xi)\big) -  \tr(V^{-1}A^+_{\rm g}) + 2 \Big]\\&
 \qquad\qquad\qquad\qquad\qquad\qquad +\frac{1}{4}\big( (\xi, g)^2 + (\xi, Jg)^2 \big) +  \frac{1}{2}\big(\xi, (P_g+P_{Jg})(\mathbb{1}+V^{-1})(\beta-\xi) \big)  \Bigg).
\end{split}
\end{equation}

With a little more rewriting, we find 
\begin{align}\label{eq:WignerDispApp}
W^{\pm}_{\xi}(\beta)=  &\frac{W_{G}(\beta - \xi)}{\tr\big( (V + \norm{\xi}^2P_{\xi} \pm \mathbb{1})(P_{g}+P_{J\!g}) \big)} \\&\times\Bigg( \norm{(P_g+P_{J\!g})(\mathbb{1}\pm V^{-1})  (\beta-\xi)}^2 +  2\big(\xi, (P_g+P_{J\!g})(\mathbb{1}\pm V^{-1}) (\beta-\xi) \big)\nonumber\\&\qquad\qquad+ \tr\big((P_g +P_{J\!g})(\norm{\xi}^2P_{\xi} - V^{-1} \mp \mathbb{1} )\big) \Bigg)\nonumber.
\end{align}

\end{document}